# Vortex shedding and heat transfer from a heated circular cylinder in Bingham plastic fluids

Sai Peng[1,2,3], Xiang Li[2,5], Li Yu[4], Xiaoru Zhuang[6] and Peng Yu[2,5*]

[1]School of Mathematics and Computational Science, Xiangtan University, Xiangtan, 411105, China

[2]Guangdong Provincial Key Laboratory of Turbulence Research and Applications, Department of Mechanics and Aerospace Engineering, Southern University of Science and Technology, Shenzhen, 518055, China

[3]National Center for Applied Mathematics in Hunan, Xiangtan, 411105, China

[4]School of Civil Engineering and Architecture, Southwest University of Science and Technology, Mianyang, 621010, China

[5]Center for Complex Flows and Soft Matter Research, Southern University of Science and Technology, Shenzhen, 518055, China

[6]School of Mechanical and Electrical Engineering, Shenzhen Polytechnic University, Shenzhen, 518055, China

**Abstract**

The present study numerically investigates the vortex shedding and heat transfer characteristics of a heated circular cylinder immersed in Bingham plastic fluids. The effects of three parameters, i.e., (i) plastic Reynolds number ($10 \leq Re \leq 180$), (ii) Prandtl number ($1 \leq Pr \leq 100$), and (iii) the Bingham number ($0 \leq Bn \leq 10^4$), are evaluated. The Navier-Stokes and energy equations for flow and heat transfer are adopted, along with the incorporation of the Papanastasiou regularization to address the discontinuous-viscosity characteristics of Bingham plastic fluids. To illustrate the impact of fluid yield stress on the flow structure, the study provides comprehensive insights into flow transition, streamlines, shear rate and velocity distributions, the morphology of yielded/unyielded regions, and the drag coefficient ($C_d$). Additionally, the temperature distribution, the local Nusselt number ($\overline{Nu_{local}}$) along the cylinder, and the average Nusselt number on the cylinder ($\overline{Nu}$) are analyzed. The results indicate that the flow transition of Bingham fluids over a circular cylinder is dependent on external disturbances, exhibiting subcritical bifurcation behavior. This leads to abrupt jumps in the $\overline{C_d}$ - *Bn* curve and the $\overline{Nu}$ - *Bn* curve near the critical Bingham number $Bn_c$. Furthermore, the heat transfer performance is contingent upon the different distribution of shear strain rate in the boundary layer across various *Bn* ranges. It is observed that $\overline{Nu}$ and *Bn* fits well with the Carreau-Yasuda-like non-Newtonian viscosity model. This investigation enhances the understanding of the vortex shedding and heat transfer behaviors in Bingham plastic fluids.

*Corresponding author (Yu P.): yup6@sustech.edu.cn

**Keywords:** Bingham plastic fluid, Vortex shedding, Heat transfer

## 1. Introduction

Viscoplastic fluids are common non-Newtonian fluids encountered in our daily products such as toothpaste and paint, as well as in various industrial applications including food processing and cosmetics[1]. Fig. 1(a) illustrates the shear stress-shear rate curve for a viscoplastic fluid, particularly a Bingham plastic fluid. Notably, this curve does not pass through the origin; instead, it intersects with the shear stress ($\tau$) axis at a point where $\tau_0 > 0$. This indicates that the minimum shear stress must exceed this critical yield value $\tau_0$ to initiate liquid flow. When shear stress is below $\tau_0$, the fluid resists flow and exhibits only elastic deformation, a behavior referred to as plastic flow.

In recent decades, the flow and heat transfer characteristics of viscoplastic fluids have garnered significant attention due to their dual nature. These fluids exhibit fluid-like behavior above the yield stress $\tau_0$ and solid-like behavior below it. From an engineering perspective, this dual nature results in the formation of yielded (fluid-like) and unyielded (solid-like) subdomains within a given flow configuration[2,3]. For instance, in the case of the flow around a cylinder, as shown in Fig. 1(b), the unyielded region of a viscoplastic fluid is manifested in the upper, lower, left and right sides of the cylinder, as well as in the surrounding outer basin area[4]. This dual-basin zone hinders both mass and heat transfer between the internal yielded region and the external unyielded region. Consequently, this system faces not only challenges related to slow and difficult mixing but also significant obstacles in convective heat transfer[5,6].

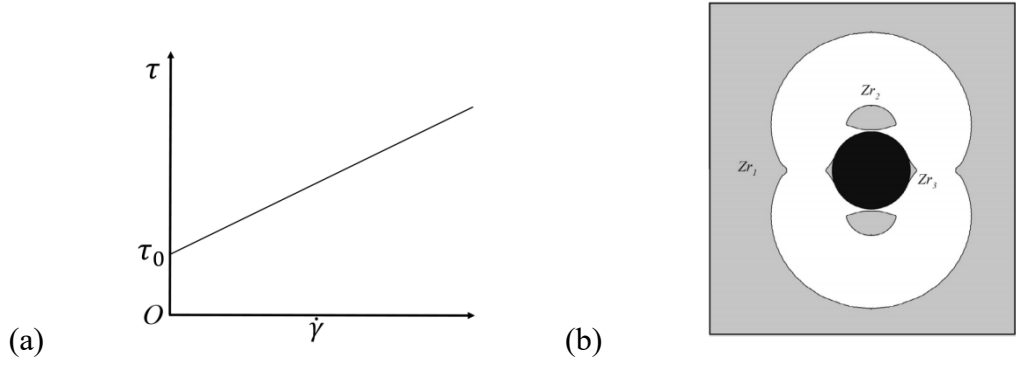

(a)      (b)

Fig. 1. (a) The relationship between shear stress ($\tau$) and shear rate ($\dot{\gamma}$) of a viscoplastic fluid. (b) The yielded (white) and unyielded (gray) regions of a viscoplastic fluid flow over a circular cylinder (the incoming flow is from left to right) at a vanishing Reynolds number, adapted from Nirmalkar & Chhabra[4].

The Bingham or Herschel-Bulkley constitutive relation is commonly employed in numerical simulations of viscoplastic fluid flows, with the plastic strength typically represented by a dimensionless Bingham number ($Bn$). Magnin and coworkers[7-9] conducted a study on the creeping flow ($Re \ll 1$) around an unconfined cylinder and the laminar Poiseuille flow past a circular cylinder confined in a plane channel. Their study focused on exploring the drag exerted on the cylinder in a viscoplastic fluid and identifying the yielded and unyielded regions. Nirmalkar and Chhabra[4] performed a numerical analysis of the heat and momentum transfer characteristics for a heated cylinder immersed in a Bingham plastic fluid, examining a range of plastic Reynolds number $1 \leq Re \leq 40$, Prandtl number $1 \leq Pr \leq 100$, and Bingham number $0 \leq Bn \leq 104$. Their comprehensive study of the flow behavior included an analysis of streamlines, yielded/unyielded region, velocity distribution, and resistance coefficient, leading to the conclusion that the presence of the yield stress inhibited the flow. Based on the extensive data, they derived a relationship for the Nusselt number ($Nu$) in the viscoelastic flow around a cylinder, expressed as $Nu \sim Re^{*1/3}Pr^{*1/3}$, where $Re^* = Re/(1+Bn)$ and $Pr^* = Pr(1+Bn)$.

Other researchers, e.g., Nirmalkar et al.[10], Thumati[11], Patel and Chhabra[12], Tiwari and Chhabra[13], Gupta and Chhabra[14], have investigated flow and heat transfer over different geometries such as a sphere, an elliptical cylinder, and a semi-circular cylinder in viscoplastic fluids. These studies have predominantly focused on steady-state conditions. However, flow over a cylinder in a viscoplastic fluid may undergo complex transitions among different flow modes, similar to those observed in Newtonian fluids, including the appearance of downstream recirculation wake and the onset of vortex shedding.

Mossaz et al.[15] investigated the recirculation wake and vortex shedding behind a cylinder in a Herschel-Bulkley fluid through numerical simulations. By regularizing the Herschel-Bulkley constitutive equation using the Papanastasiou model, they analyzed the influence of $Bn$ number ($0 \leq Bn \leq 10$) on the flow patterns, particularly concerning the unyielded region. Their results indicated that both the critical Reynolds numbers and Strouhal numbers for the onset of the recirculation wake and vortex shedding increased with $Bn$. The two critical Reynolds numbers relationships were approximately expressed as $Re_{c1} = 48.3Bn + 7$ and $Re_{c2} = 45.8Bn + 47$. Moreover, they reported that an increase in inertial force (higher $Re$) tended to expand the yielded regions spatially; however, this trend was countered by the effect of yield stress (higher $Bn$). This interplay not only suppressed fluid detachment from the cylinder surface but also inhibited vortex shedding.

Since previous studies on the viscoplastic flow around a cylinder have primarily focused on steady-state conditions[4,7-9], the underlying mechanisms on the unsteady flow phenomena, such as vortex shedding, remain inadequately understood[15]. For instance, finite disturbance often occurs during the transition in non-Newtonian flow[16,17], yet it is unclear whether the initial transition to vortex shedding in a viscoplastic fluid shows disturbance dependence (subcritical behavior). Heat transfer in viscoplastic fluids is critical in various industrial situations, such as food processing[18-20]. Efficient control of heat transfer in plastic fluids is vital to ensuring products quality and safety[21]. Furthermore, in Newtonian fluids, exceeding a critical transition threshold in the Reynolds number can significantly enhance convective heat transfer in unsteady flow[22]. Therefore, it is essential to examine heat transfer in viscoplastic fluid around a cylinder when the Reynolds number exceeds a corresponding critical threshold.

In this study, we focus on numerically solving the momentum and energy equations governing the unsteady thermal flow of a Bingham plastic fluid around a heated circular cylinder. The parameters considered include the plastic Reynolds number ranging from 10 to 180, the Prandtl number from 1 to 100, and the Bingham number from 0 to $10^4$. The extensive results on the flow and thermal fields (including streamlines, isotherms, and the morphology of yielded/unyielded regions) and the global parameters (such as the drag coefficient and the Nusselt number) are presented. These results elucidate the influence of the plastic Reynolds number, the Prandtl number, and the Bingham number on the dynamics and heat transfer characteristics of a Bingham plastic flow past a cylinder.

## 2. Mathematical model and governing equations

### 2.1. Problem description

Consider the scenario of an incompressible and unsteady flow of a Bingham plastic fluid, characterized by a uniform incoming velocity $\mathbf{u} = (U_\infty, 0)$ and temperature $T_0$, over a heated circular cylinder with a diameter $D$, as depicted in Fig. 2(a). The surface of the cylinder is maintained at a constant temperature $T_w$, which is higher than $T_0$. The cylinder is positioned at the center of the computational domain, with the distances between the cylinder center and the inlet and outlet boundaries set to $L_u = 25D$ and $L_d = 75D$, respectively. The lateral width of the computational domain is denoted as $H = 50D$, ensuring a blockage ratio ($BR = D/H$) of 2%.

The block-structured mesh has been generated for the present computational domain using the commercial software ANSYS ICEM. The region surrounding the cylinder is discretized using an O-type mesh, as depicted in Fig. 2(b). Other regions within the computational domain are discretized using multiple blocks of rectangular meshes, with a denser mesh allocation near the cylinder and a coarser mesh allocation near the domain boundaries. The O-type mesh consists of 400 grid points uniformly distributed along the cylinder perimeter and 121 grid points stretched exponentially in the radial direction to ensure a high-resolution mesh near the cylinder surface. In this study, the size of the first cell adjacent to the cylinder surface in the radial direction is set to $0.0025D$. In the $x$ direction, 501 grid points (for $L_d$) are distributed unevenly in the downstream region, while 101 grid points (for $L_u$) are positioned in the upstream region. To accurately capture the temperature gradient near the cylinder surface at a high Prandtl number, the mesh is further intensified locally adjacent to the cylinder surface, as shown in Fig. 2(c). In this case, the thickness of the grid closest to the cylinder wall is set at $0.000625D$. The total number of meshes for the computational domain is 209,600.

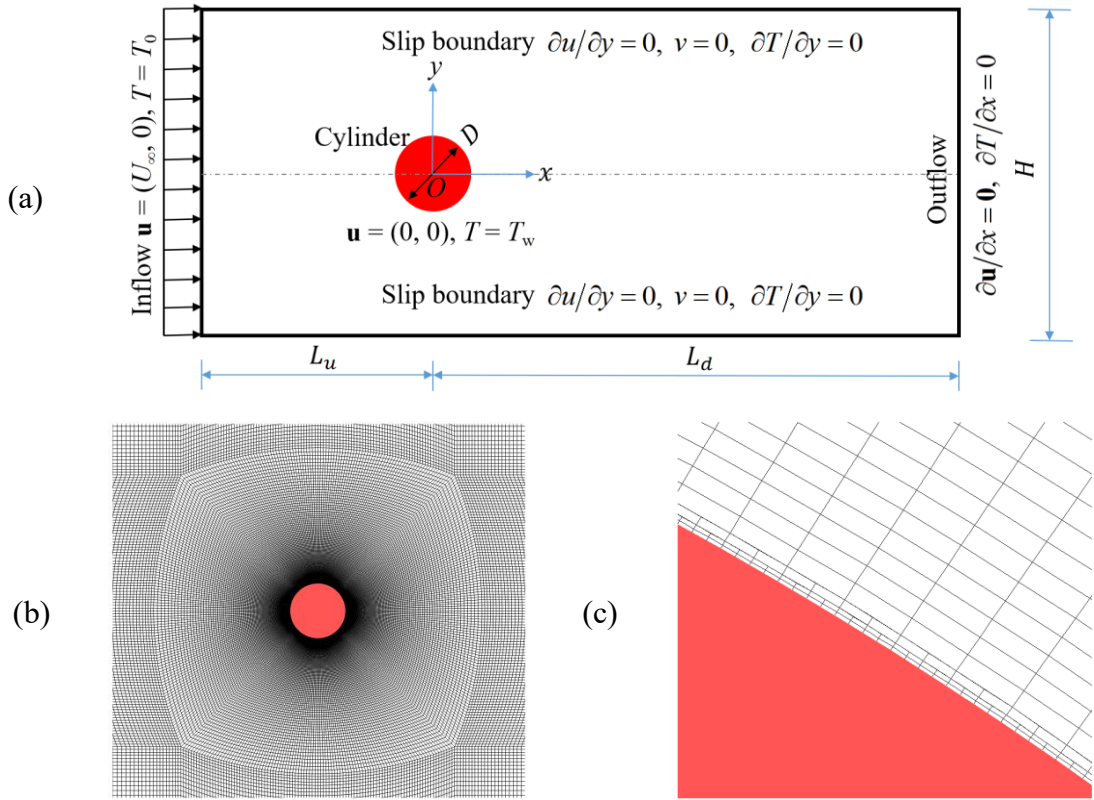

Fig. 2. (a) Schematic of the computational domain. (b) The mesh around the cylinder. (c) The enlarged view of the mesh near the cylinder surface.

## 2.2. Governing equations

In this simulation, the thermo-physical properties of the fluid, i.e., thermal conductivity $k$, heat capacity $C_p$, plastic viscosity $\mu_B$, yield stress $\tau_0$, and density $\rho$, are assumed to be independent of temperature. Additionally, the effect of viscous dissipation is considered negligible. While these assumptions allow for the decoupling of the velocity and temperature fields, they also restrict the applicability of the present results to the situations where the temperature difference $\Delta T = T_w - T_o$ is sufficiently small. Under these assumptions, the equations of continuity, momentum, and thermal energy can be expressed as follows,

$$\nabla \cdot \mathbf{u} = 0, \tag{1}$$

$$\rho \frac{\partial \mathbf{u}}{\partial t} + \rho \mathbf{u} \cdot \nabla \mathbf{u} = -\nabla p + \nabla \cdot \boldsymbol{\tau}. \tag{2}$$

$$\rho C_p \left( \frac{\partial T}{\partial t} + \mathbf{u} \cdot \nabla T \right) = k \nabla^2 T. \tag{3}$$

where $\mathbf{u}$ is the velocity vector filed, $p$ is the pressure field, $T$ is the temperature field, and $\boldsymbol{\tau}$ is the extra stress tensor. For a Bingham plastic fluid, $\boldsymbol{\tau}$ could be written as,

$$\boldsymbol{\tau} = \mu_B \dot{\gamma} + \frac{\tau_0}{\sqrt{\Pi_{\dot{\gamma}}}} \dot{\gamma}, \text{ if } \Pi_{\dot{\gamma}} > \tau_0, \tag{4}$$

$$\dot{\gamma} = 0, \text{ if } \Pi_{\dot{\gamma}} < \tau_0, \tag{5}$$

where $\dot{\gamma}$ is the rate of deformation tensor given as follows:

$$\dot{\gamma} = \nabla \mathbf{u} + \nabla \mathbf{u}^T. \tag{6}$$

The magnitudes of these two tensors are frequently required in the calculation of yielded/unyielded regions and given as follows:

$$|\dot{\gamma}| = \sqrt{\Pi_{\dot{\gamma}}}, \ |\tau| = \sqrt{\Pi_\tau}, \tag{7}$$

where

$$\Pi_{\dot{\gamma}} = \text{tr}(\dot{\gamma}^2), \ \Pi_\tau = \text{tr}(\tau^2). \tag{8}$$

To avoid the discontinuity in the Bingham plastic constitutive equation, the equation is regularized using the Papanastasiou model[18], which has been adopted in many recent studies:[4,7-12]

$$\boldsymbol{\tau} = \mu_B \dot{\gamma} + \frac{\tau_0 \cdot \left[1 - \exp\left(-M \sqrt{\Pi_{\dot{\gamma}}}\right)\right]}{\sqrt{\Pi_{\dot{\gamma}}}} \dot{\gamma}, \tag{9}$$

where $M$ denotes the regularization parameter.

The governing and constitutive equations noted above are non-dimensionalized using $D$ and $U_\infty$ as the length and velocity scales, respectively. These scales, along with other properties, can be used to derive additional scales, e.g., $\mu_B D/U_\infty$ for stress components, $U_\infty/D$ for the rate of deformation tensor, $\rho U_\infty^2$ for pressure, among others. This non-dimensionalization indicates that the velocity field is influenced by two dimensionless parameters: the plastic Reynolds number and the Bingham number. Furthermore, the temperature field exhibits an additional dependence on the Prandtl number. These dimensionless parameters are defined here as follows:

**Bingham number**

$$Bn = \frac{\tau_0 D}{\mu_B U_\infty}. \tag{10}$$

Note that $Bn \to 0$ and $Bn \to \infty$ correspond to the Newtonian flow and the fully plastic flow, respectively.

**Plastic Reynolds number**

$$Re = \frac{\rho D U_\infty}{\mu_B}. \tag{11}$$

**Prandtl number**

$$Pr = \frac{C_p \mu_B}{k}. \tag{12}$$

The drag and lift force coefficients on the cylinder are calculated as:

$$C_d = \frac{2\oiint [-p\mathbf{n} + \boldsymbol{\tau} \cdot \mathbf{n}]_x \, dS}{\rho U_\infty^2 D}, \tag{13}$$

$$C_l = \frac{2\oiint [-p\mathbf{n} + \boldsymbol{\tau} \cdot \mathbf{n}]_y \, dS}{\rho U_\infty^2 D}, \tag{14}$$

where $\mathbf{n}$ is a unit vector in the outward normal direction and $dS$ is the infinitesimal element of area on the cylinder surface.

For unsteady flow, the flow data such as lift and drag are collected over more than 10 cycles to calculate the corresponding statistical data once the flow reaches statistical stationary state. The time-averaged drag coefficient and the root mean square lift coefficient are calculated as follows,

$$\overline{C_d} = \frac{1}{m}\sum_{i=1}^{m} C_d(t_i), \quad t_i > t_o, \tag{15}$$

$$C_{lrms} = \sqrt{\frac{1}{m-1}\sum_{i=1}^{m}\left[C_l(t_i)\right]^2}, \quad t_i > t_o, \tag{16}$$

where $t_o$ represents the time instant when the flow reaches statistical stationary state, and *m* is the total number of statistical moments. In this paper, adding a bar above a variable denotes the time-averaged value of this variable.

The Strouhal number is introduced to quantify the frequency (*f*) of vortex shedding and defined as follows,

$$St = \frac{fD}{U_\infty}. \tag{17}$$

*f* is obtained by fast Fourier transform (FFT) of the time series of the lift coefficient when $t > t_o$.

The local Nusselt number ($Nu_{local}$) is introduced to evaluate the heat transfer performance on the cylindrical surface, which is defined as,

$$Nu_{local} = \frac{D}{T_o - T_w}\cdot\frac{\partial T}{\partial n}. \tag{18}$$

The time-averaged local Nusselt number ($\overline{Nu_{local}}$) is calculated as,

$$\overline{Nu_{local}} = \frac{1}{m}\sum_{i=1}^{m} Nu_{local}(t_i), \quad t_i > t_o, \tag{19}$$

The overall Nusselt number (*Nu*) along the cylinder wall is adopted to evaluate the overall heat dissipation effect on the cylindrical surface, which is expressed as,

$$Nu = \frac{\oint Nu_{local}\cdot dS}{\pi D}. \tag{20}$$

For unsteady case, the time-averaged Nusselt number ($\overline{Nu}$) along the cylinder wall is written as,

$$\overline{Nu} = \frac{1}{m}\sum_{i=1}^{m} Nu(t_i), \quad t_i > t_o, \tag{21}$$

## 2.3. Numerical method

To solve the aforementioned equations numerically, the commercial finite-volume solver ANSYS FLUENT is employed. Comprehensive information regarding the computational methods utilized within this software can be found in various sources[23-26]. Here, we provide only a brief overview. The quadratic upstream interpolation for convective kinematics and second-order implicit discretization schemes are adopted to discretize the spatial and temporal domains, respectively. At the inlet boundary, a uniform streamwise velocity $\mathbf{u} = (U_\infty, 0)$ is imposed, along with a fixed temperature $T_o$. The transverse boundaries of the simulation domain are treated as the symmetric boundary condition. Pressure is set to 0 at the outlet boundary. The velocity on the cylinder surface adheres to a no-slip condition, that is, $\mathbf{u} = (0, 0)$, and a fixed temperature $T_w$ ($T_w > T_o$) is set on the cylinder surface. Initially, an unsteady simulation is conducted to determine whether the flow is steady or unsteady. If the flow is checked to be steady, a steady state calculation is subsequently employed. Conversely, if the flow identified as unsteady, the simulation result over ten cycles is used once the flow reaches statistical stationary state.

Table 1. Comparison of the statistical parameters for various time steps at (*Re, Bn*) = (100, 5)

| $\frac{\Delta t \cdot U_\infty}{D}$ | $\overline{C_d}$ | $C_{lmax}$ | $St$ | $\overline{Nu}$ | | |
|---|---|---|---|---|---|---|
| | | | | $Pr = 1$ | $Pr = 10$ | $Pr = 100$ |
| 0.01 | 1.3658 | 0.1864 | 0.1546 | 5.7318 | 13.5437 | 30.2411 |
| 0.005 | 1.3636 | 0.1805 | 0.1553 | 5.7294 | 13.5397 | 30.2287 |
| 0.0025 | 1.3624 | 0.1775 | 0.1557 | 5.7279 | 13.5353 | 30.2226 |
| 0.00125 | 1.3618 | 0.1762 | 0.1558 | 5.7270 | 13.5325 | 30.2189 |

The independence of the time step is verified and summarized in Table 1. The case selected for this analysis corresponds to (*Re, Bn*) = (100, 5). The time step (Δ*t*) investigated are $0.01\frac{U_\infty}{D}$, $0.005\frac{U_\infty}{D}$, $0.0025\frac{U_\infty}{D}$, and $0.00125\frac{U_\infty}{D}$. The results for $\frac{\Delta t \cdot U_\infty}{D} = 0.00125$ are close to those for $\frac{\Delta t \cdot U_\infty}{D} = 0.0025$. For example, $\overline{C_d}$ is 1.3618 when $\frac{\Delta t \cdot U_\infty}{D} = 0.00125$ while $\overline{C_d}$ is 1.3624 when $\frac{\Delta t \cdot U_\infty}{D} = 0.0025$. In order to balance computational efficiency with the required accuracy, the time step in this study is set as $\frac{\Delta t \cdot U_\infty}{D} = 0.0025$.

In this study, the Papanastasiou model (Eq. 9) is employed to regularise the Bingham constitutive equation. As the parameter $M$ increases, the error associated with the regularization model decreases. However, a larger value of $M$ can adversely affect the stability of numerical calculation. The influence of $M$ on the statistical parameters is investigated and summarized in Table 2, with the case corresponding to $(Re, Bn) = (100, 5)$ selected for analysis. $\frac{M \cdot U_\infty}{D}$ is set as $10^3$, $10^4$, $10^5$, and $10^6$. The results for $\frac{M \cdot U_\infty}{D} = 10^5$ are close to those of $\frac{M \cdot U_\infty}{D} = 10^6$. For example, $\overline{C_d}$ is 1.3624 when $\frac{M \cdot U_\infty}{D} = 10^5$ while $\overline{C_d}$ is 1.3631 when $\frac{M \cdot U_\infty}{D} = 10^6$. To balance the stability and the accuracy of numerical simulation at the same time, $M$ is set as $\frac{M \cdot U_\infty}{D} = 10^5$ in this study. Nirmalkar & Chhabra[4] also selected $\frac{M \cdot U_\infty}{D} = 10^5$ while Mossaz et al.[15] utilized $\frac{M \cdot U_\infty}{D} = 10^6$.

Table 2. Comparison of the statistical parameters for various $M$ at $(Re, Bn) = (100, 5)$

| $\frac{M \cdot U_\infty}{D}$ | $\overline{C_d}$ | $C_{lmax}$ | $St$ | $\overline{Nu}$ | | |
|---|---|---|---|---|---|---|
| | | | | $Pr = 1$ | $Pr = 10$ | $Pr = 100$ |
| $10^3$ | 1.3443 | 0.1712 | 0.1540 | 5.7045 | 13.4732 | 30.0996 |
| $10^4$ | 1.3578 | 0.1756 | 0.1554 | 5.7217 | 13.5190 | 30.1910 |
| $10^5$ | 1.3624 | 0.1775 | 0.1557 | 5.7279 | 13.5353 | 30.2226 |
| $10^6$ | 1.3631 | 0.1777 | 0.1557 | 5.7286 | 13.5372 | 30.2261 |

## 3. Results and Discussion

### 3.1. Flow and heat transfer behavior for a Newtonian fluid

Newtonian fluid flow over a circular cylinder has been extensively investigated through experiments and numerical simulations. Once $Re$ surpasses a critical Reynolds number $Re_c$, the flow transitions from a steady to an unsteady state. In our simulation, the predicted $Re_c$ is 46.1, accompanied by the corresponding critical Strouhal number ($St_c$) of 0.1168. The values of $Re_c$ and $St_c$ obtained from literature are summarized in Table 3. It can be seen that $Re_c$ ranges from 46 and 48, while $St_c$ falls between 0.1168 and 0.132. Our results exhibit strong agreement with the data reported in the previous studies[27-32].

Table 3. $Re_c$ and $St_c$ for a Newtonian fluid flow over a cylinder.

| Source | $Re_c$ | $St_c$ |
|---|---|---|
| Present | 46.1 | 0.1168 |
| Williamson (1989)[27] | 47.9 | 0.122 |
| Norberg (1994, 2001)[28,29] | 47.4 | 0.122 |
| Sivakumar et al. (2006)[30] | 46-47 | 0.1179 |
| Kumar and Mittal (2006)[31] | 46.8 | 0.1168 |
| Morzynski et al. (1999)[32] | 47 | 0.132 |

The variation of the time-averaged drag coefficient ($\overline{C_d}$) with $Re$ obtained from our simulation is compared with the finding of Sen et al.[33], Qu et al.[34], and Park et al.[35], as shown in Fig. 3(a). It is observed that $\overline{C_d}$ decreases with increasing $Re$ within the range of $10 < Re < 180$. However, for the range of $50 < Re < 180$, the decline rate of $\overline{C_d}$ becomes very small. Our simulation results align closely with those reported in literature[33-35]. The flow transition from steady to unsteady state is regarded as a supercritical Hopf bifurcation[36]. In the unsteady state, $C_{lrms}$ is not equal to zero. The relationship between $C_{lrms}$ and $Re$ is shown in Fig. 3(b), alongside the results from Qu et al.[34] and Park et al.[35]. The present results are consistent with the published data. As illustrated in Figs. 3(b) and 3(c), the relationship between $C_{lrms}$ and $Re$ in the present study can be expressed by the following equation,

$$C_{lrms} = \frac{(Re-Re_c)^{0.6554}}{56.9401} \text{ or } \log C_{lrms} = 0.6554\log(Re - Re_c) - 4.042. \quad (22)$$

The linear relationship between $\log C_{lrms}$ and $\log(Re - Re_c)$ is confirmed by Fig. 3(c).

The correlation between *St* and *Re* is compared with the available data, as shown in Fig. 3(d). Williamson[27] proposed the following empirical formula,

$$St = -\frac{3.3265}{Re} + 0.1816 + 0.00016 Re. \tag{23}$$

A similar equation was provided by Norberg[28], which reads as follows,

$$St = -\frac{3.458}{Re} + 0.1835 + 0.000151 Re. \tag{24}$$

The comparison in Fig. 3(d) indicates that our simulation results coincide well with Williamson[27] and Norberg[28].

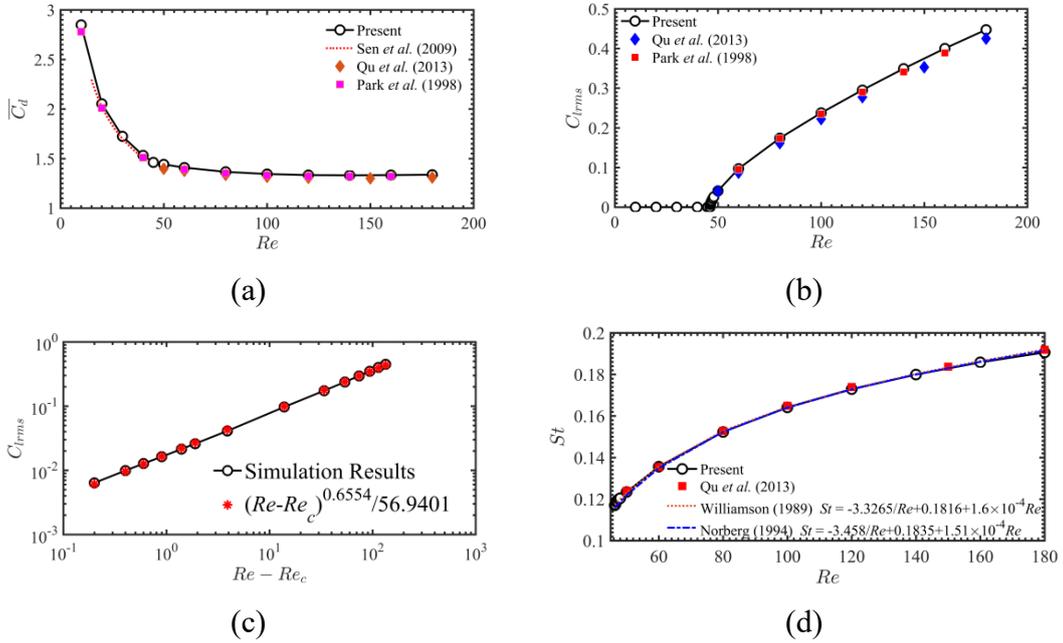

Fig. 3. (a) $\overline{C_d}$~*Re*, (b) $C_{lrms}$~*Re*, (c) $C_{lrm}$ ~ (*Re* - *Re*c), and (d) *St*~*Re* curves for a Newtonian fluid flow over a cylinder.

The relationship between the time-averaged Nusselt number ($\overline{Nu}$) and *Re*, together with the published data, are illustrated in Fig. 4. Kramers[37], Salimipour[38], and Sarkar *et al.*[39] provided the following empirical formulas for $\overline{Nu}$ and *Re*, respectively:

$$\overline{Nu} = 0.42 Pr^{0.20} + 0.57 Pr^{0.33} Re^{0.50}, (5 \leq Re \leq 1000), \tag{25}$$

$$\overline{Nu} = 0.42 Pr^{0.20} + 0.57 Pr^{0.33} Re^{0.50}, \tag{26}$$

$$\overline{Nu} = 0.459 Pr^{0.373} Re^{0.548}. (80 \leq Re \leq 180 \text{ and } 0.7 \leq Pr \leq 100). \tag{27}$$

Our numerical results exhibit strong agreement with those reported in literature[37-39], as depicted in Fig. 4. In our simulation, the relationship between $\overline{Nu}$ and (*Pr*, *Re*) satisfies the following piecewise function, with the relative error less than 5%,

$$\overline{Nu} = \begin{cases} 0.6039 Pr^{1/3} Re^{1/2}, & (10 \leq Re \leq 45) \\ 0.5111 Pr^{0.358} Re^{0.532}, & (50 \leq Re \leq 180) \end{cases}. \quad (28)$$

Our numerical results indicate that it is difficult to represent $\overline{Nu}$ as a single continuous power-law relation with respect to $Pr$ and $Re$. Instead, a discontinuity is observed when flow transitions from a steady to an unsteady state. This discontinuity may be attributed to the flow fluctuation that contributes to the additional heat transfer enhancement when $Re$ exceeds $Re_c$.

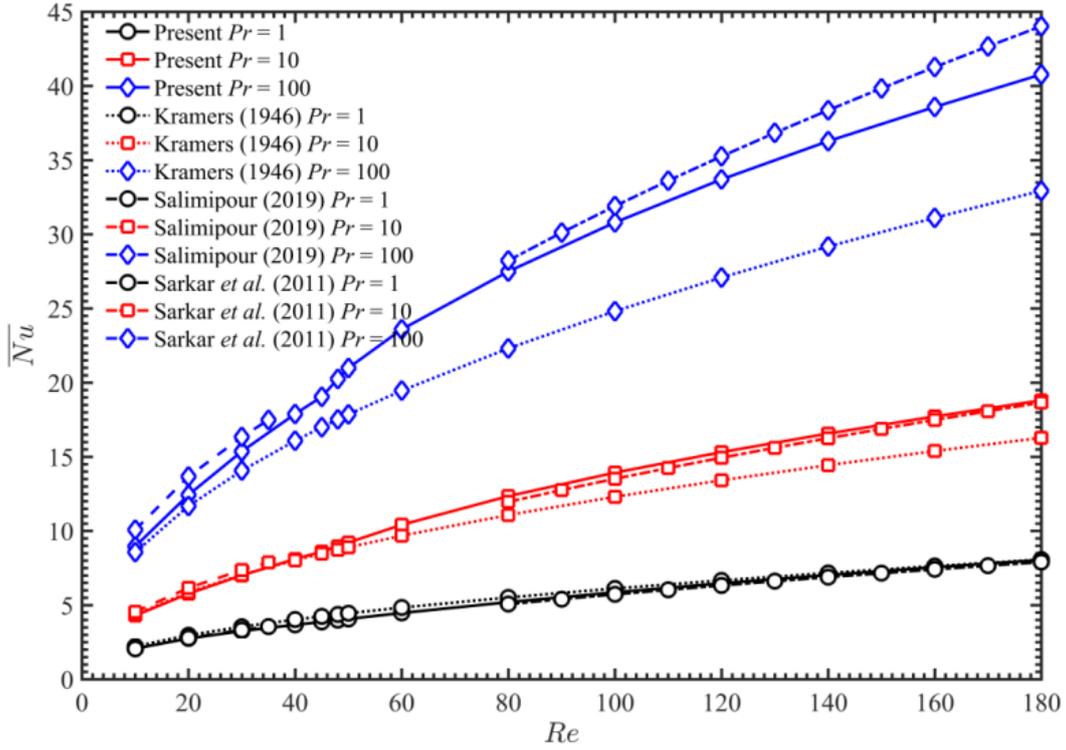

Fig. 4. Variation of $\overline{Nu}$ with $Re$ at different $Pr$ for a Newtonian fluid flow over a cylinder.

The validations conducted above serve to affirm that the current methodology can proficiently simulate the thermal flow of a Newtonian fluid around a circular cylinder with a high degree of accuracy.

## 3.2. Flow and heat transfer behavior for a Bingham plastic fluid

### 3.2.1. Flow feature

For the steady flow over the cylinder, the recirculation wake may disappear once $Bn$ exceeds another critical value, denoted as $Bn_c^*$. For example, at $Re = 40$ and $Bn = 0$, a pair of symmetrical recirculation wake appears behind the cylinder, as illustrated in Fig. 5(a). When $Bn$ is increased to 1, the recirculation wake noticeably decreases in size. Furthermore, the recirculation wake completely vanishes at $Bn = 2$. It should be note that the exact value of $Bn_c^*$ is not considered in this paper. At higher $Re$, $Bn_c^*$ also increases. For example, the recirculation wake disappears for $Bn$ between 5 and 20 at $Re = 100$ and $Re = 180$, as shown in Figs. 5(b) and 5(c). The instantaneous streamlines in Fig. 5 indicate that the influence of $Bn$ on the wake dynamics is more pronounced at a high $Re$, underscoring the complex interplay between these parameters in the flow field around the cylinder.

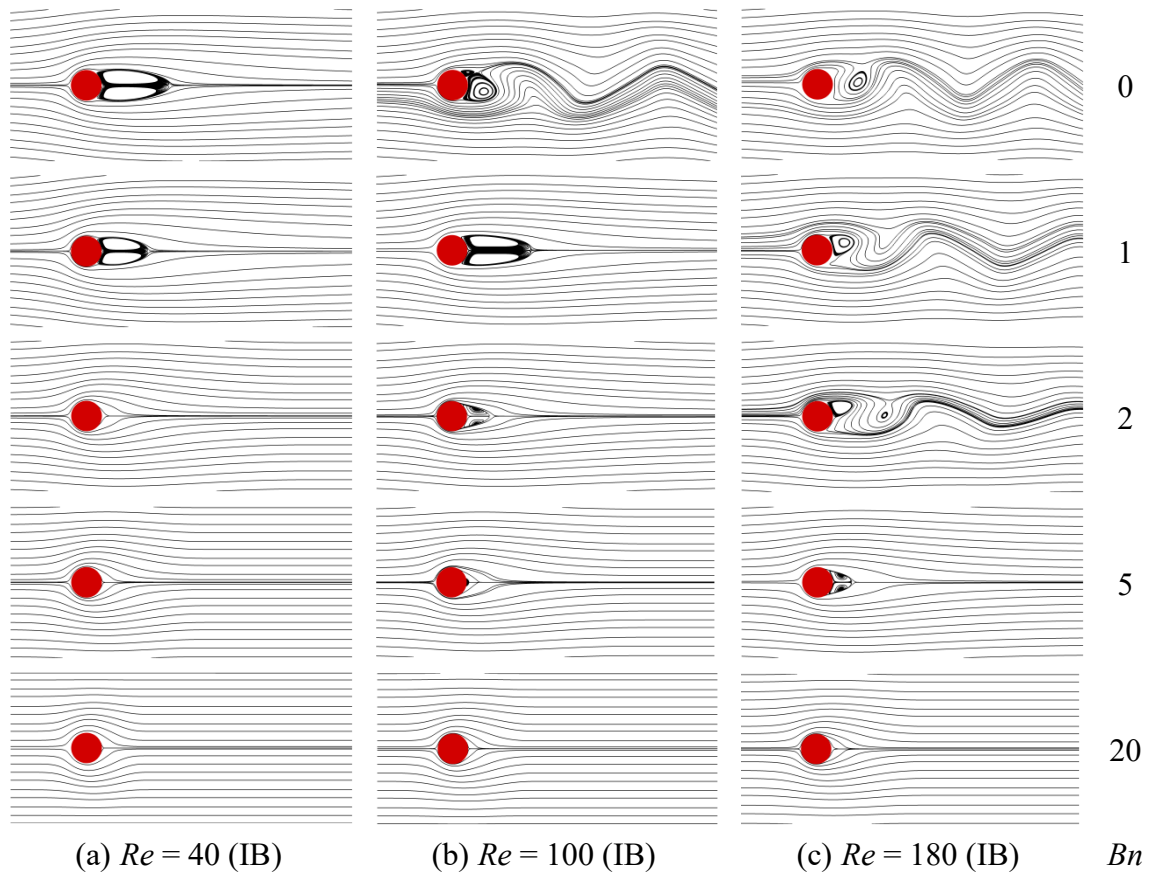

(a) $Re = 40$ (IB)     (b) $Re = 100$ (IB)     (c) $Re = 180$ (IB)     $Bn$

Fig. 5. Instantaneous streamlines for different $Re$ and $Bn$.

For a transition from the steady to unsteady state in a Newtonian flow over a cylinder, the critical Reynolds number $Re_c$ is not influenced by initial disturbance, which indicates a supercritical bifurcation. Conversely if $Re_c$ shows sensitive to initial disturbance, the bifurcation is termed as a subcritical one. In this study, the effect of the intensity of the initial disturbance on the onset of vortex shedding for a Bingham plastic fluid over a cylinder is examined. Mossaz et al.[15] introduced a large initial disturbance by artificially rotating the cylinder at a constant rate until flow oscillations emerged (thus triggering an instability). Another alternative way to adjusting the intensity of the disturbance is to use the numerical result of the current state to initialize the next simulation while gradually increasing or decreasing the control parameter $Bn$ at a fixed $Re$[17]. For example, in the case of $(Re, Bn) = (100, 0.1)$, the final flow field obtained for $Re = 100$ in a Newtonian fluid ($Bn = 0$) is adopted to initialize the simulation. This process is referred to as increasing $Bn$ process and is denoted as IB. On the other hand, if the simulation case of $Re = 100$ and $Bn = 2$ is initialized by using the final flow field from $Re = 100$ and $Bn = 3$, this process is termed as the decreasing $Bn$ process and denoted as DB. Generally, the intensity of the disturbance of the DB process is lower than that in the IB process for a specified $Bn$.

The instantaneous streamlines at various $Re$ and $Bn$ are illustrated in Fig. 5. The spatial-temporal instability of the flow field can be assessed by observing whether the upper and lower symmetry of the streamlines behind the cylinder is preserved. An increase in $Bn$ leads to flow stabilization. For instance, at $Re = 100$, vortex shedding exists behind the cylinder in a Newtonian fluid, whereas, in a Bingham plastic fluid with $Bn = 1$, vortex shedding disappears completely. Similarly, at $Re = 180$, vortex shedding disappears for $Bn$ ranging over 2 to 5. The critical Bingham number ($Bn_c$) for the suppression of vortex shedding at various $Re$ is summarized in Fig. 6. However, our simulations shows that $Bn_c$ for the IB and DB processes may not be identical, and are denoted as $Bn_{cI}$ and $Bn_{cD}$, respectively. The stability of the flow of a Bingham plastic fluid around a cylinder is particularly influenced by the initial disturbance when $Re$ exceeds 60, suggesting a subcritical bifurcation in the onset of vortex shedding. The difference between $Bn_{cI}$ and $Bn_{cD}$ increases with $Re$. Mossaz et al.[15] observed an approximate linear relationship between $Bn_{cI}$ and $Re$, described as follows,

$$Bn_{cI} = 0.0218\,Re - 1.0262, (0 < Bn < 10). \tag{29}$$

For the IB process, our simulations provide a similar linear relationship between $Bn_{cI}$ and $Re$, which reads,

$$Bn_{cI} = 0.0201\,Re - 0.9993. \tag{30}$$

The fitting coefficients in this study are very close to those in Mossaz et al.[15]

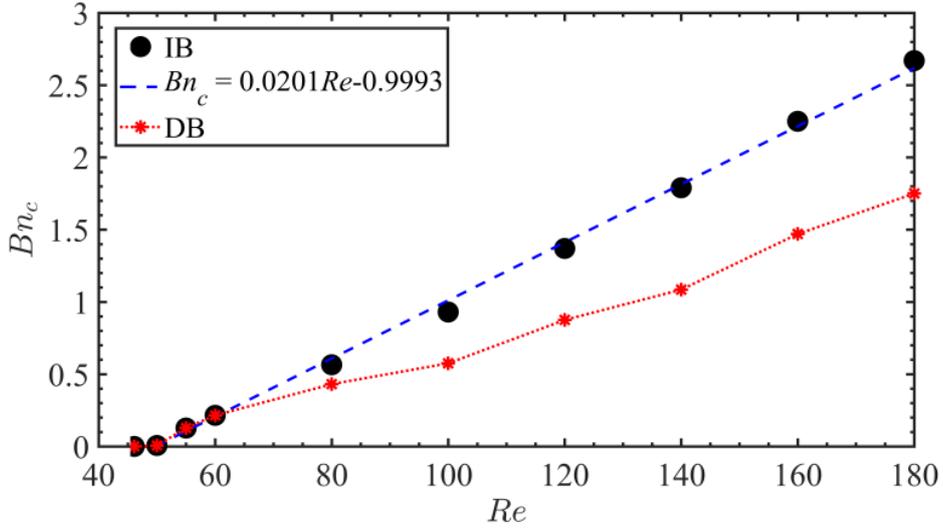

Fig. 6. Variation of $Bn_c$ with $Re$ in a Bingham plastic fluid.

The disappearances of downstream recirculation wake and vortex shedding can be attributed to the elevated shear viscosity in a Bingham plastic fluid as $Bn$ increases. This increase in shear viscosity is a consequence of the elastic solid-like behavior when the shear stress is below the yield stress. A bi-viscous criterion is applied to determine whether the flow yields. Specifically, the flow yields when $\mu/\mu_B < 10^5$.[4]

The yielded and unyielded regions for various $Re$ and $Bn$ are depicted in Fig. 7. The yielded region is predominantly observed in the vicinity surrounding the cylinder, excluding the front, rear, top, and bottom sides of the cylinder when $Bn$ is sufficiently high, such as $Bn = 10^4$. This behavior is similar to that observed at $Re = 0$ as shown in Fig. 1(b). According to the flow yield characteristics, the flow field may be divided into six regions: the yield region (denoted by the white color), one unyielding region $Zr_1$, two unyielding regions $Zr_2$ that are located near the lateral sides of the cylinder characterized by a narrow gap between $Zr_2$ and the cylinder surface), and two unyielding regions $Zr_3$ attached to the front and rear of the cylinder, as shown in Fig. 1(b).

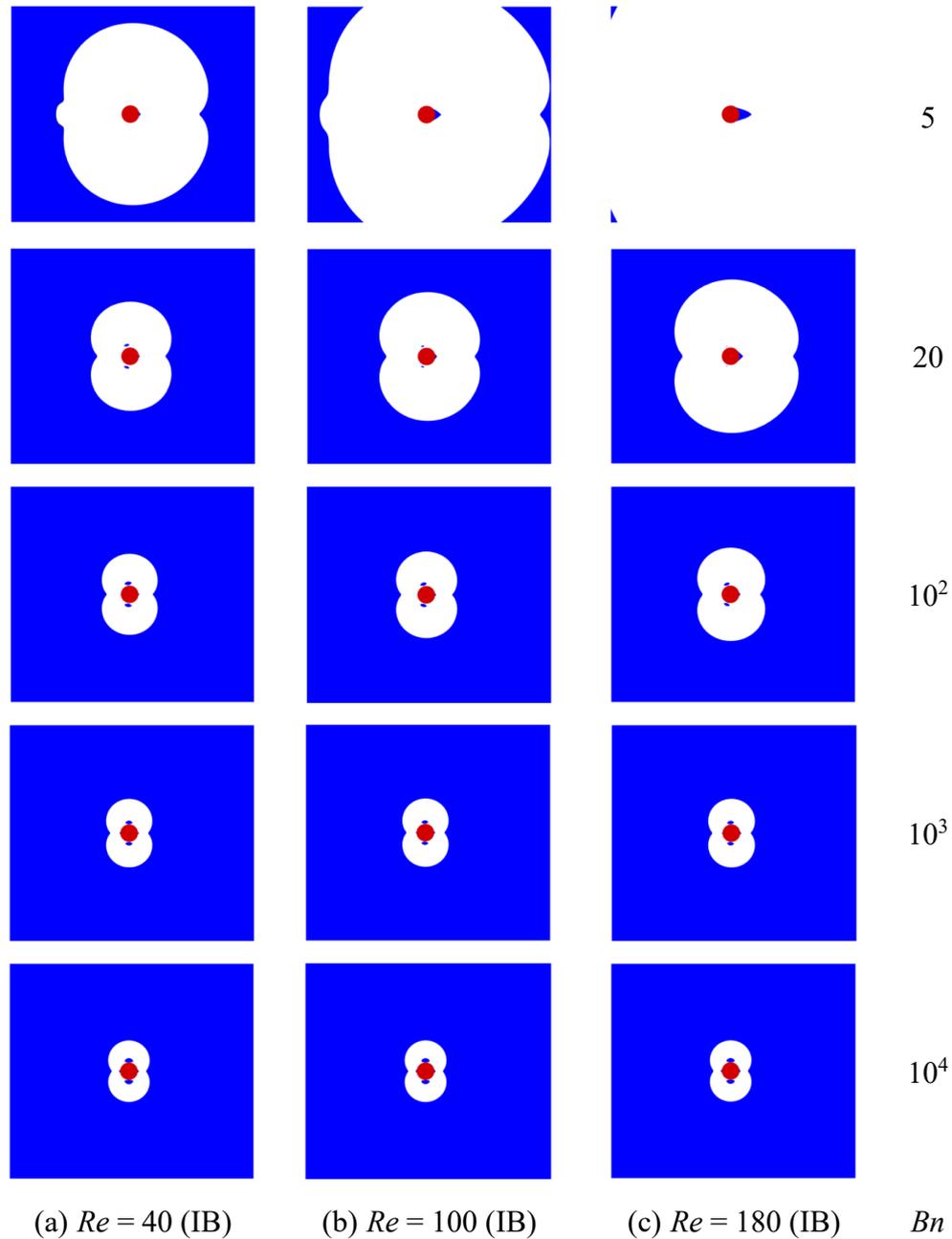

(a) $Re = 40$ (IB)    (b) $Re = 100$ (IB)    (c) $Re = 180$ (IB)    $Bn$

**Fig. 7.** Influence of $Re$ and $Bn$ on the morphology of the yielded (white) and unyielded (blue) regions. The regions are determined based on the time-averaged flow field.

The presence of $Zr_3$ is associated with the velocity stagnation points located upstream and downstream of the cylinder. At a fixed $Bn$, the yielded region (shown in white in Fig. 7) expands as $Re$ increases. And the trend is more pronounced at lower $Bn$. Conversely, with a fixed $Re$, an increase in $Bn$ leads to a reduction in the size of the yielded region. The high viscosity downstream of the cylinder contributes to suppressing flow instability[40]. Consequently, as shown in Fig. 6, the flow transition in the flow is delayed. The behaviors of $Zr_2$ and $Zr_3$ exhibit distinct characteristics. At a fixed $Bn$, $Zr_3$ expands while $Zr_2$ shrinks as $Re$ increases. This phenomenon is also more pronounced at a lower $Bn$. Conversely, at a fixed $Re$, $Zr_2$ occurs and expands both upstream and downstream, while $Zr_3$ shrinks as $Bn$ increases, with the upstream and downstream symmetry of $Zr_3$ becoming more pronounced.

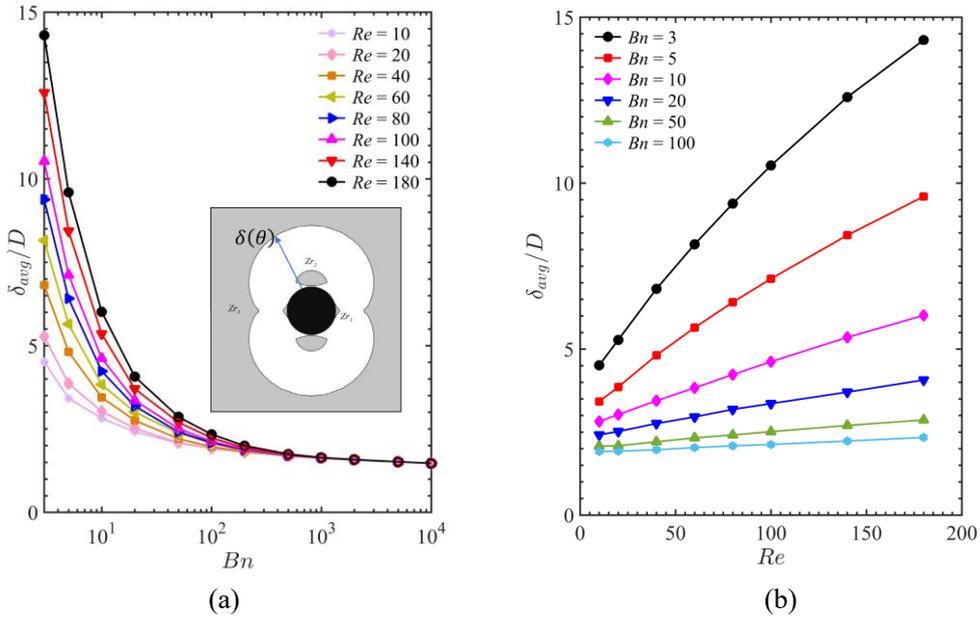

(a)          (b)

**Fig. 8.** Influence of $Re$ and $Bn$ on the average value ($\delta_{avg}$) of the dividing line thickness $\delta(\theta)$ between the unyielding outer region and the yielded inner region. The gray region ($Zr_1$, $Zr_2$ and $Zr_3$) in the illustration represent unyielding regions, while the white region represents the yield region.

The radial distance between the inner boundary of the $Zr_1$ (represented by the contour line for $\mu/\mu_B = 10^5$) and the cylinder surface $\delta(\theta)$, as illustrated in the inset of Fig. 8, is calculated. The average distance $\delta_{avg}$ (averaging $\delta(\theta)$ along the circumference) is then computed, which provides a rough estimate of the size of the yielded region. The variation of $\delta_{avg}/D$ with $Re$ and $Bn$ is presented in Fig. 8. At a fixed $Re$, $\delta_{avg}/D$ decreases with increasing $Bn$, indicating that a larger $Bn$ corresponds to a smaller the yielded region. An increase in $Bn$ signifies that the flow is more difficult to yield as a whole. Within the range of parameters investigated, at a fixed *Bn*, $\delta_{avg}/D$ increases with *Re*, suggesting that the inertial force enhances the overall yielding for flow around the cylinder. It is worth pointing out that the effect of $Zr_2$ and $Zr_3$ on the size of the yielded region is not discussed here.

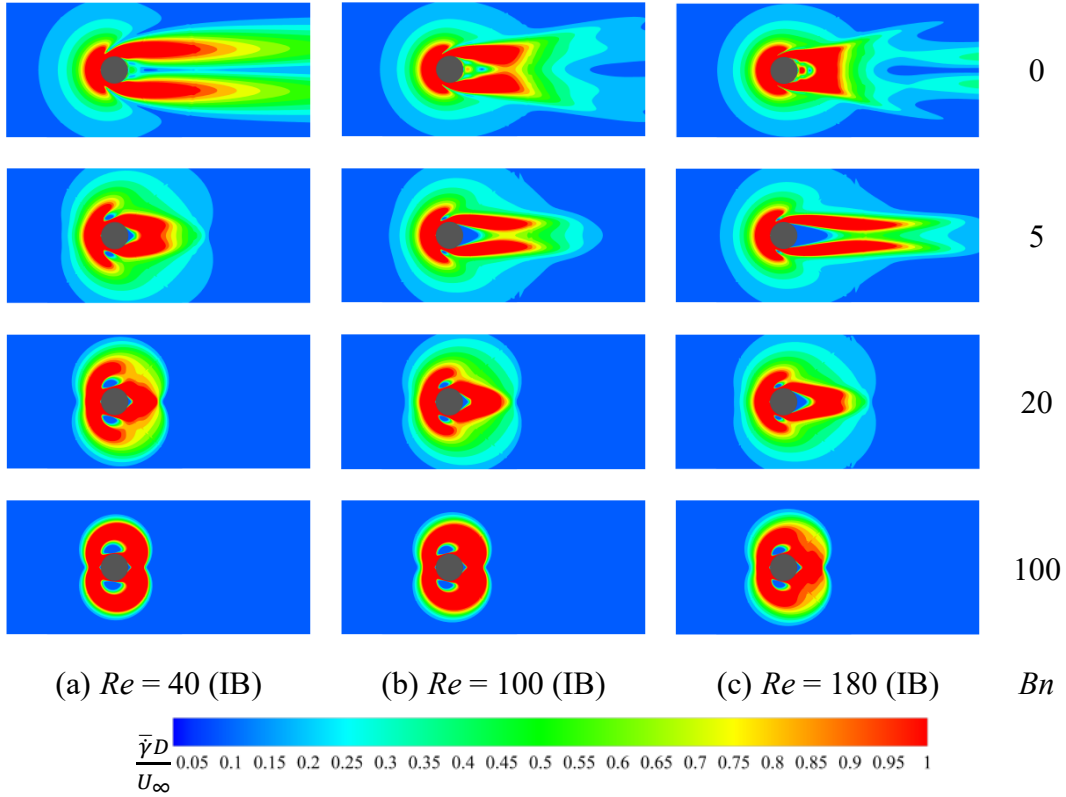

(a) *Re* = 40 (IB)   (b) *Re* = 100 (IB)   (c) *Re* = 180 (IB)   *Bn*

$\frac{\bar{\dot{\gamma}} D}{U_\infty}$  0.05 0.1 0.15 0.2 0.25 0.3 0.35 0.4 0.45 0.5 0.55 0.6 0.65 0.7 0.75 0.8 0.85 0.9 0.95 1

**Fig. 9.** The distributions of normalized time-averaged shear strain rate at different *Bn* with *Re* = (a) 40, (b) 100, and (c) 180 for the IB process.

When the shear stress and the shear strain rate exceed their respective thresholds, flow yielding occurs. The distributions of the normalized time-averaged shear strain rate ($\bar{\dot{\gamma}} D/U_\infty$) for various *Re* and *Bn* are shown in Fig. 9. The region with high shear strain rate is primarily located around the cylinder. As *Bn* increases, the symmetry of upstream and downstream of the cylinder in the region with high shear strain rate becomes more pronounced. Moreover, the high shear strain rate region that extends downstream of the cylinder gradually diminishes in size. Nevertheless, three small regions with relatively low shear strain rate are observed surrounding the cylinder: one $Zr_3$ region located directly behind and attached to the cylinder, and two $Zr_2$ regions located above and below the cylinder. With an increase in $Bn$, the downstream region with low shear strain rate diminishes, while the upper and lower regions with low shear strain rate become wider.

The yielded behavior in the regions of $Zr_2$ and $Zr_3$ are directly or indirectly associated with the boundary layer near the cylinder surface. Thus, the velocity profile in the vicinity of the cylinder is analyzed in this section. The $u$-velocity profiles along the horizontal center line of the cylinder (in the $x$-direction) for various $Re$ and $Bn$ are depicted in Fig. 10(i). In the case of unsteady flow, the time-averaged $u$-velocity ($\bar{u}$) profile is depicted.

In a Newtonian fluid, a negative velocity region is observed behind the cylinder, corresponding to the downstream recirculation wake. As $Bn$ increases, the interval for the negative velocity becomes shorter, corresponding to a shrinking recirculation wake. When $Bn$ exceeds $Bn_c^*$, the negative $u$-velocity disappears, signifying the disappearance of the recirculation wake. For example, in the cases of $Re = 180$ and $Bn = 5$ illustrated in Fig. 5, two symmetrical recirculation regions are present behind the cylinder. As $Bn$ increases, both the region with the negative velocity (as shown in Fig. 10(i)) and the recirculation wake (as shown in Fig. 5) gradually narrow and ultimately disappear. This phenomenon leads to a compact unyielded region of $Zr_3$ attached to the cylinder (as shown in Fig. 7), characterized by high shear viscosity in the wake downstream of the cylinder.

Fig. 10(ii) illustrates the $u$-velocity profiles along the vertical centerline of the cylinder (in the $y$-direction) for various $Re$ and $Bn$, with the corresponding enlarged view near the cylinder surface shown in Fig. 10(iii). At the cylinder surface ($y/D = 0.5$), the $u$-velocity equals zero. In all cases, the overall trend of $u$-velocity with respect to $y$ follows a consistent pattern. As $y$ increases, the $u$-velocity gradually increases to the maximum value ($u_{max}$), which exceeds 1. Then, as $y \to \infty$, the $u$-velocity restores to the incoming flow velocity. Specially, in a Bingham plastic fluid, $u_{max}$ is greater than that in a Newtonian fluid. At a high $Bn$, the velocity gradient near the cylinder is notably large, as shown in Fig. 10(iii).

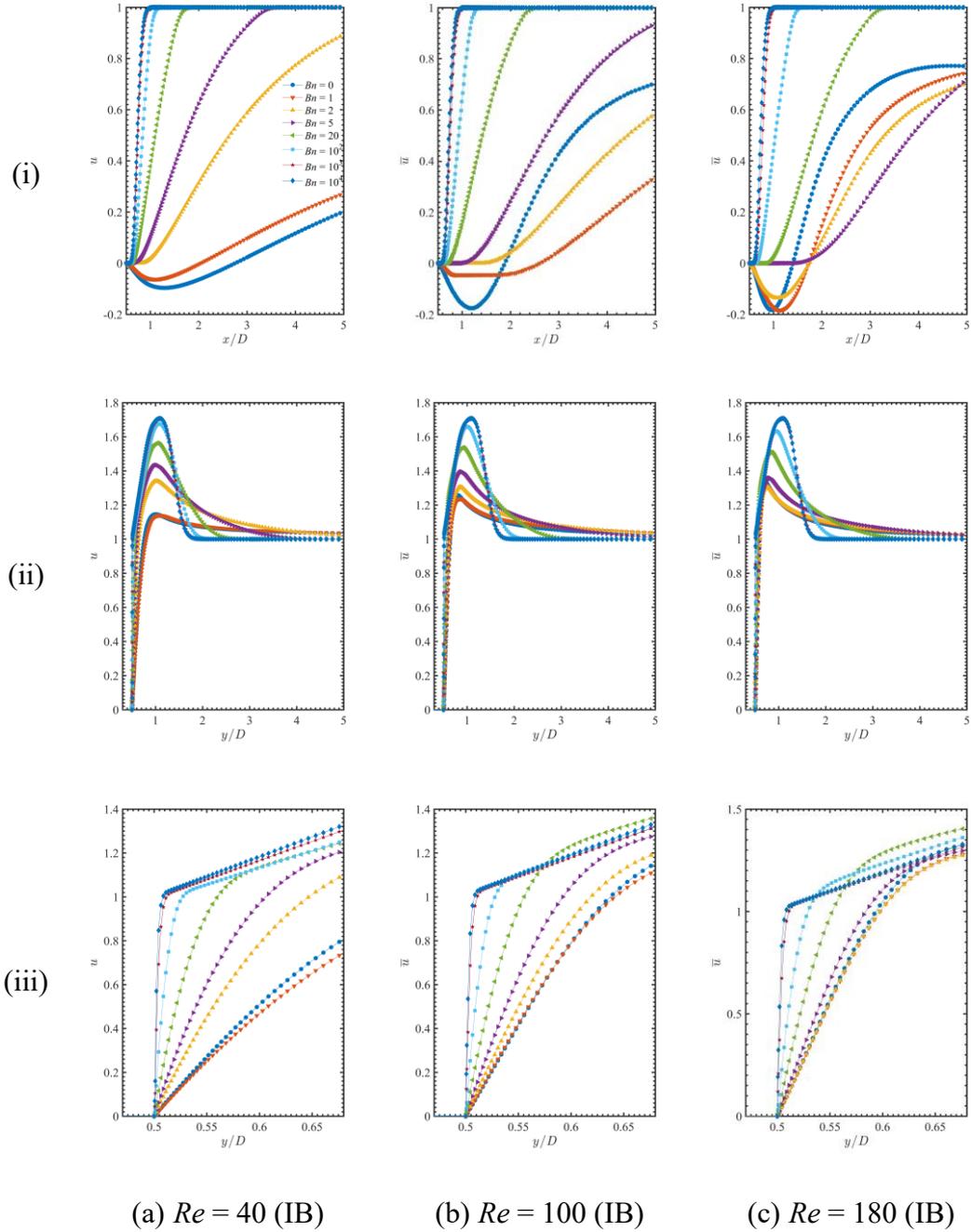

(a) *Re* = 40 (IB)　　(b) *Re* = 100 (IB)　　(c) *Re* = 180 (IB)

**Fig. 10.** The *u*-velocity profiles for different *Bn* at *Re* = (a) 40, (b) 100, and (c) 180 for the IB process. (i) and (ii) denote the *u*-velocity profiles along the horizonal (in the *x*-direction) and vertical (in the *y*-direction) center lines of the cylinder, respectively. (iii) is the enlarged view of (ii). For *Re* =100 and 180, the time-averaged *u*-velocity profiles are plotted.

The normalized shear strain rate profiles along the horizontal center line of the cylinder (in the *x*-direction) are plotted in Fig. 11(i). Only one peak exists when $Re = 40$ for all *Bn* and $Re = 100$ and 180 for high *Bn*. When $Re = 100$ or 180 and low *Bn*, two peaks exist. The first peak (the maximum shear strain rate) locates at the center of the recirculation wake due to the negative tail velocity. As *Bn* increases, the first peak is significantly reduced as shown in Fig. 11(i), due to the shrinking recirculation wake as shown in Fig. 5. The maximum shear strain rate along the horizontal center line of the cylinder (in the *x*-direction) for different *Bn* and *Re* are plotted in Fig. 12(i). Obviously, with the increase of *Bn*, the maximum shear strain rate first decreases and then increases for all *Re*. Although the yielded region behind the cylinder rear gradually expands as shown in Figs. 7 and 9, a higher velocity gradient appears near the cylinder at a high *Bn*. This indicates a more complicated flow behavior with the boundary layer.

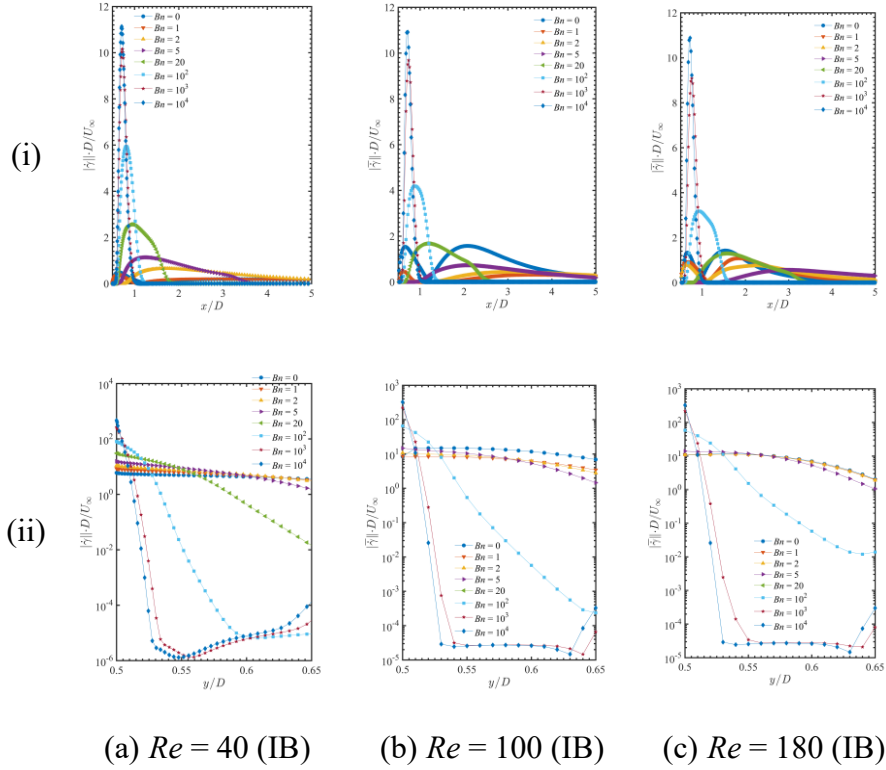

(a) *Re* = 40 (IB)   (b) *Re* = 100 (IB)   (c) *Re* = 180 (IB)

**Fig. 11.** The normalized shear strain rate profiles at different *Bn* with *Re* = (a) 40, (b) 100, and (c) 180 for the IB process. (i) and (ii) denote the shear strain rate profiles along the horizonal (in the *x*-direction) and vertical (in the *y*-direction) center lines of the cylinder, respectively. For *Re* =100 and 180, the time-averaged shear strain rate profiles are plotted.

The shear strain rate profiles along the vertical center line of the cylinder (in the *y*-direction) at different *Bn* and *Re* are plotted in Fig. 11(ii), which shows that the maximum shear strain rate occurs on the cylinder surface. Consequently, the flow attached to the cylinder is more likely to yield but the yielding region shrinks, as shown in Fig. 7. However, beyond the boundary layer, the flow gradient is smoothed out, as shown in Fig. 11(ii), resulting an unyielded region as indicated in Fig. 7. This velocity distribution near the cylinder is similar to that of the plastic channel boundary layer theory[41].

The maximum shear strain rate along the vertical center line of the cylinder for various $Bn$ and $Re$ are summarized in Fig. 12(ii). For a low $Bn$, the maximum shear strain rate of different $Bn$ slightly changes at a fixed $Re$. However, when $Bn$ exceeds a certain critical value (which increases with $Re$), the maximum shear strain rate significantly increases. For example, $\frac{\dot{\gamma}_{max} \cdot D}{U_\infty} = 325.4298$ when $Bn = 10^4$ and $\frac{\dot{\gamma}_{max} \cdot D}{U_\infty} = 8.5224$ when $Bn = 1$ for $Re = 40$.

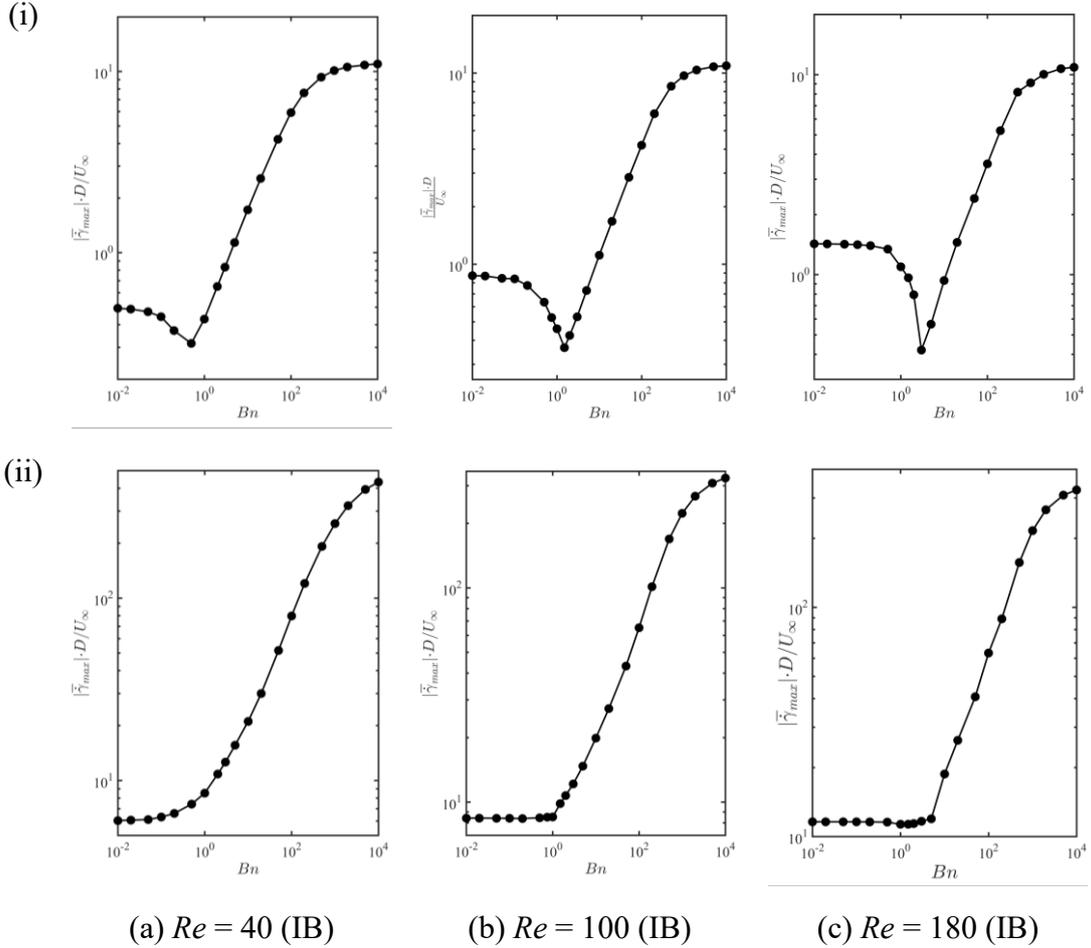

(a) $Re = 40$ (IB)  (b) $Re = 100$ (IB)  (c) $Re = 180$ (IB)

**Fig. 12.** Variation of the maximum normalized shear strain rate with $Bn$ at $Re =$ (a) 40, (b) 100, and (c) 180 for the IB process. (i) and (ii) denote the maximum normalized shear strain rate along the horizonal (in the $x$-direction) and vertical (in the $y$-direction) center lines of the cylinder, respectively. For $Re =$100 and 180, the time-averaged maximum normalized shear strain rate is plotted.

In summary, with an increase in $Bn$, the overall flow plasticization near the cylinder is obviously enhanced, leading to a narrowing of the boundary layer thickness near the cylinder surface. This reduction in boundary layer thickness is associated with a stronger velocity gradient in that region.

Once $C_d$ and $C_l$ on the cylinder have been obtained, $C_{lrms}$, $St$ and $\overline{C_d}$ can be subsequently calculated. Fig. 13 shows the variations of $C_{lrms}$ and $St$ with $Bn$ at different $Re$. $C_{lrms}$ could be used to characterize the general behavior of flow fluctuation near the cylinder wall, with the zero-value confirming the steady flow. For a fixed $Re$, an increase in $Bn$ results in a significant decrease in $C_{lrms}$, indicating a weakening of flow fluctuation. At low $Bn$, $C_{lrms}$ and $Bn$ approximately satisfy a linear relationship, as indicated by the dashed line in Fig.13(a). However, in the IB process, as $Bn$ approaches a critical value $Bn_{cI}$, $C_{lrms}$ suddenly drops from a finite value to zero, which is more obvious at a higher $Re$. For example, at $Re = 180$ (IB), $C_{lrms}$ is 0.1448 when $Bn = 2.68$ and $C_{lrms}$ is 0 when $Bn = 2.7$.

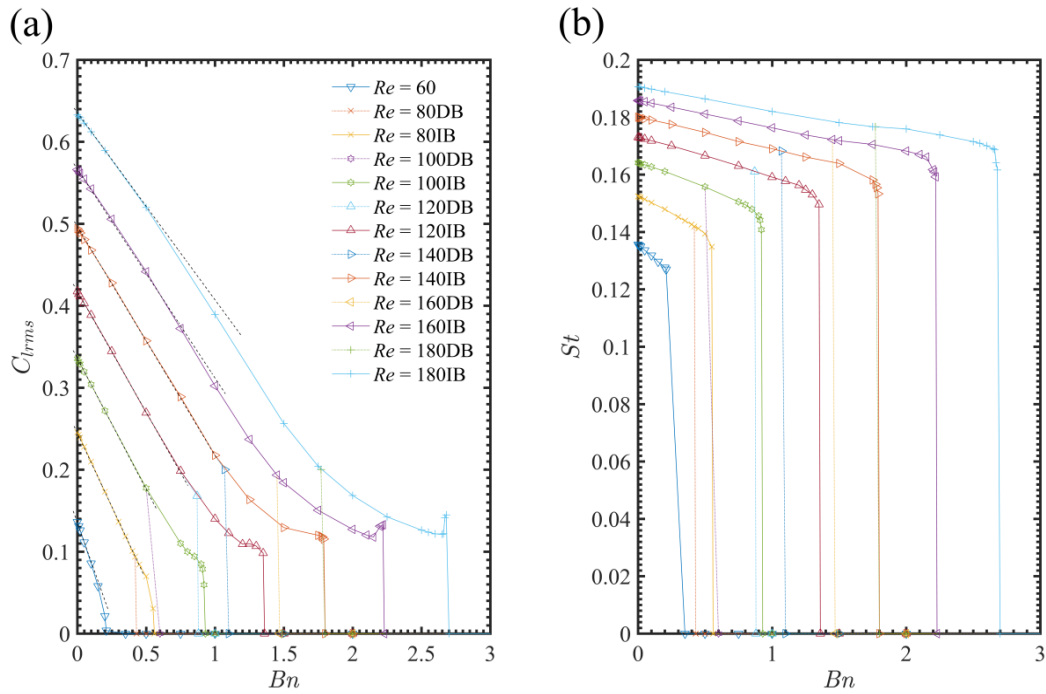

Fig. 13. Variations of (a) $C_{lrms}$ and (b) $St$ with $Bn$ at different $Re$. The dashed lines represent an approximate linear relationship.

For the DB process and $Re \geq 60$, when $Bn$ approaches another critical value $Bn_{cD}$, $C_{lrms}$ shoots up from zero to a finite value. For example, when $Re = 180$ (DB), $C_{lrms}$ is 0 when $Bn = 1.8$ and $C_{lrms}$ is 0.2002 when $Bn = 1.77$. These observations suggest that the flow behavior around a cylinder in a Bingham plastic fluid is highly disturbance-dependent within the range of $Bn_{cI}$ and $Bn_{cD}$. Additionally, Fig. 13(b) shows that $St$ decreases with increasing $Bn$, a trend that can be attributed to the enhanced shear viscosity in the wake region of a plastic Bingham fluid, as illustrated in Fig. 7. Similar to $C_{lrms}$, $St$ displays a kind of hysteresis behavior.

For the Stokes flow of a Newtonian fluid, the nonlinear terms in the governing equations can be ignored and then the Navier-Stokes equations degenerate into a series of linear equations. Theoretically, $C_d$ is inversely proportional to $Re$, and this relationship can be expressed as,[42]

$$C_d = X/Re, (Re \ll 1), \quad (31)$$

where $X$ is a constant and related to the computational spatial domain and the boundary conditions. Eq. (31) can be rewritten in the double logarithmic coordinate system,

$$\log(C_d) = -\log(Re) + \log(X), (Re \ll 1), \quad (32)$$

where $C_d$ and $Re$ satisfy a linear relationship. Lamb[42] calculated $X = 12.5538$ for an infinite domain.

As shown in Fig. 7, the shear viscosity is obviously higher than the plastic viscosity $\mu_B$. Taking $U_\infty/D$ as the characteristic strain rate, an effective shear viscosity could be defined as follows,

$$\mu_{eff} = \mu_B + \frac{\tau_0 D}{U_\infty} = \mu_B \cdot (1+Bn). \quad (33)$$

A modified Reynolds number can then be defined as,

$$Re^* = \frac{\rho D U_\infty}{\mu_{eff}} = \frac{Re}{1+Bn}. \quad (34)$$

As indicated by Eq. (34), a higher $Bn$ corresponds to a lower $Re^*$.

For the steady flow of Bingham plastic fluids, the comparison of $C_d$ between our results and those reported by Nirmalkar & Chhabra[4] and Mossaz et al.[15] is presented in

Table 4. The relative error between our result and that of Mossaz *et al.*[15] is less than 5%, and the corresponding error between our result and that of Nirmalkar & Chhabra[4] is less than 10%, which indicate good agreement.

**Table 4.** Drag coefficient ($C_d$) for a cylinder in Bingham plastic fluids at different *Re* and *Bn*.

| *Re* | *Bn* | Nirmalkar & Chhabra[4] | Mossaz *et al.*[15] | Present |
|---|---|---|---|---|
| 10 | 1 | 6.7955 | 6.8994 | 6.8306 |
| 10 | 5 | 20.458 | 19.405 | 19.075 |
| 10 | 10 | 34.788 | 33.105 | 32.929 |
| 20 | 1 | 3.8340 | 3.9749 | 3.9330 |
| 20 | 5 | 10.571 | 10.192 | 10.025 |
| 20 | 10 | 17.214 | 16.996 | 16.904 |
| 20 | $10^4$ | 12166.1 | - | 12049.0 |
| 40 | 1 | 2.3532 | 2.4262 | 2.4191 |
| 40 | 5 | 5.6276 | 5.5597 | 5.4732 |
| 40 | 10 | 9.4278 | 8.9614 | 8.9197 |

Fig. 14(a) illustrates the variation of $\overline{C_d}$ with $Re^*$ for various *Re* in the double logarithmic coordinate system. A linear relationship between $\log(\overline{C_d})$ and $\log(Re^*)$ is observed when $Re^* \ll 1$. This observation is consistent with the simulation results of Nirmalkar & Chhabra[4] for *Re* ranging from 1 to 40. Moreover, our results indicate that the linear relationship between $\log(\overline{C_d})$ and $\log(Re^*)$ for $Re^* \ll 1$ is still valid for *Re* more than 40. Our simulation suggests that *X* is approximately equal to 24.84, which gives,

$$\overline{C_d} = 24.84/Re^*. \qquad (35)$$

Note that *X* = 24.84 is very close to 24.75 reported by Nirmalkar & Chhabra[4], but is obviously higher than 12.5538 in a Newtonian fluid reported by Lamb[42].

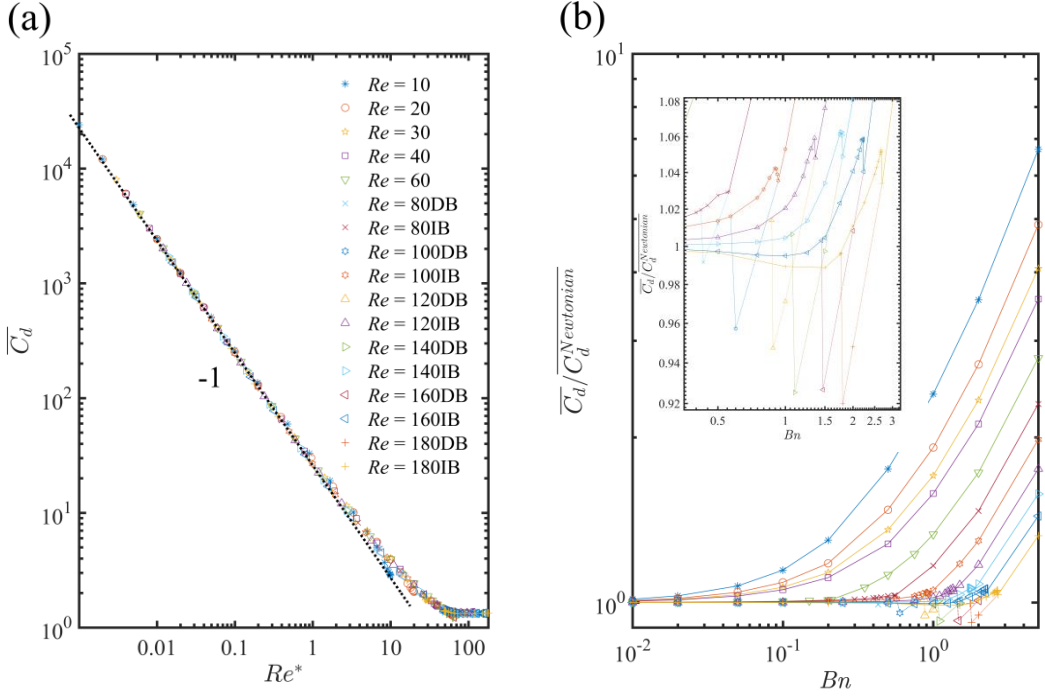

Fig. 14. (a) Variation of $\overline{C_d}$ with $Re^*$ at different $Re$ and (b) variation of $\overline{C_d}/\overline{C_d^{Newtonian}}$ with $Bn$ at different $Re$.

When $Re^*$ is below 0.5, the relative error of Eq. (35) is less than 10%. However, when $Re^*$ exceeds 1, the linear relationship between $\log(\overline{C_d})$ and $\log(Re^*)$ no longer holds, as shown in Fig. 14(a). To gain a deeper understanding on the behavior of $\overline{C_d}$ at small $Bn$, Fig. 14(b) depicts the variation of $\overline{C_d}/\overline{C_d^{Newtonian}}$ with $Bn$ at various $Re$. Generally, $\overline{C_d}/\overline{C_d^{Newtonian}}$ increases with $Bn$ for all $Re$. The curves for $Re \leq 60$ exhibit a gradual growth, whereas for $Re \geq 80$, the variation of $\overline{C_d}/\overline{C_d^{Newtonian}}$ with $Bn$ becomes sharp near $Bn_{cI}$ and $Bn_{cD}$. For example, in the IB process at $Re = 180$, $\overline{C_d}/\overline{C_d^{Newtonian}}$ drops suddenly from 1.052 to 1.034 when $Bn$ is slightly increased from 2.68 to 2.7. This abrupt reduction in $\overline{C_d}/\overline{C_d^{Newtonian}}$ is attributed to the transition from the unsteady ($C_{lrms} \neq 0$) to steady flow ($C_{lrms} = 0$), as shown in Fig. 13(a). Conversely, for the DB process at $Re = 180$, $\overline{C_d}/\overline{C_d^{Newtonian}}$ shoots up from 0.9198 to 0.9963 when $Bn$ is slightly decreased from 1.8 to 1.77. The significant increase in $\overline{C_d}/\overline{C_d^{Newtonian}}$ results from the transition from the steady ($C_{lrms} = 0$) to unsteady flow ($C_{lrms} \neq 0$), as shown in Fig. 13(a). With the exception of the behaviors observed near $Bn_{cI}$ and $Bn_{cD}$, the variation of $\overline{C_d}/\overline{C_d^{Newtonian}}$ with $Bn$ remains smooth across the present range investigated.

### 3.2.2. Heat transfer feature

The heat transfer behavior of a Bingham fluid flow past a circular cylinder is discussed in this section. We will discuss the influence of $Re$, $Bn$, and $Pr$ on the dimensionless temperature ($\frac{T-T_0}{T_w-T_0}$) field individually, as demonstrated in Figs. 15-17.

At a low $Re$, such as $Re = 40$, a thick thermal boundary layer with a high temperature region around the cylinder is observed in Newtonian fluid, as illustrated in Fig. 15(a). In this scenario, thermal conduction is still significant. As $Re$ increases, the influence of convection intensifies, resulting in a reduction in the thickness of thermal boundary layer. Special to Bingham plastic fluids, the thermal boundary layer thickness also decreases with an increase in $Bn$. The evidence for this fact is that the high temperature zone behind the cylinder at a fixed $Re$ converges towards the flow field center line with increasing $Bn$. This observation coincides well with the findings reported by Nirmalkar & Chhabra[4].

Heat is transferred from the cylinder surface to the downstream wake of the cylinder along the flow direction. When the flow becomes unsteady, vortex shedding occurs behind the cylinder, accompanied by the release of hot 'blobs'. This behavior is exemplified in the flow and temperature fields for the case of ($Re$, $Bn$, $Pr$) = (180, 2, 1) as shown in Figs. 5(c) and 15(c). For the IB process, when $Bn$ exceeds $Bn_{cI}$, the flow becomes steady and the shedding of hot 'blobs' ceases. This stabilization is observed in the flow and temperature fields for the case of ($Re$, $Bn$, $Pr$) = (180, 5, 1), as shown in Figs. 5(c) and 15(c).

As $Bn$ exceeds a threshold, the thermal boundary layer around the cylinder shrinks, causing a more concentrated high-temperature region near the horizontal center line behind the cylinder. This phenomenon is evident in Fig. 15 for $Bn = 5$ and $Bn = 20$. As discussed in Figs. 9-12, a higher $Bn$ results in a thinner momentum boundary layer thickness and a sharper shear strain rate in the momentum boundary layer. This thinner momentum boundary layer consequently leads to a thinner temperature boundary layer thickness.

As $Pr$ increases, the high-temperature region behind the cylinder becomes more concentrated along the center line of the flow field, which could be seen from Figs. 15(a), 16(a) and 17(a) at a fixed $Re = 40$ for various $Pr = 1$, 10, and 100. The influence of $Pr$ is relatively straightforward to comprehend. $Pr$ represents the relative ratio between the thickness of momentum boundary layer and the thickness of thermal boundary layer. Thus, a higher $Pr$ indicates a thinner thermal boundary layer when $Re$ is fixed.

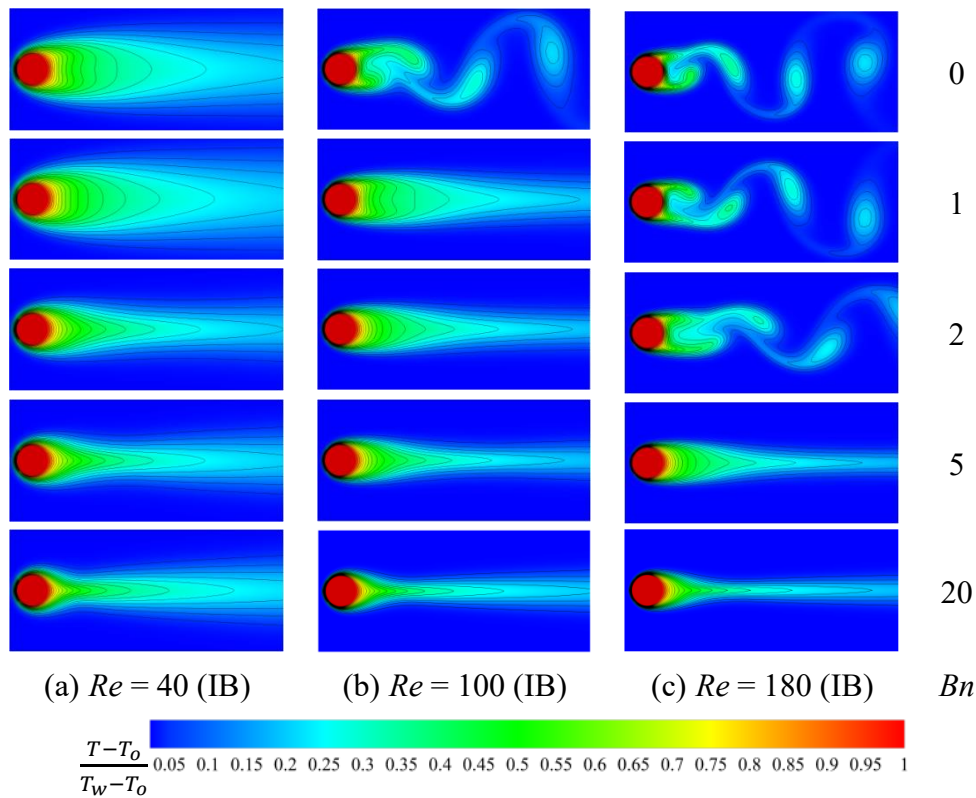

Fig. 15. Effect of $Bn$ on the dimensionless temperature distributions at $Re =$ (a) 40, (b) 100, and (c) 180 for the IB process at a fixed $Pr = 1$. The legends in Figs. 16 and 17 are the same as that of Fig. 15.

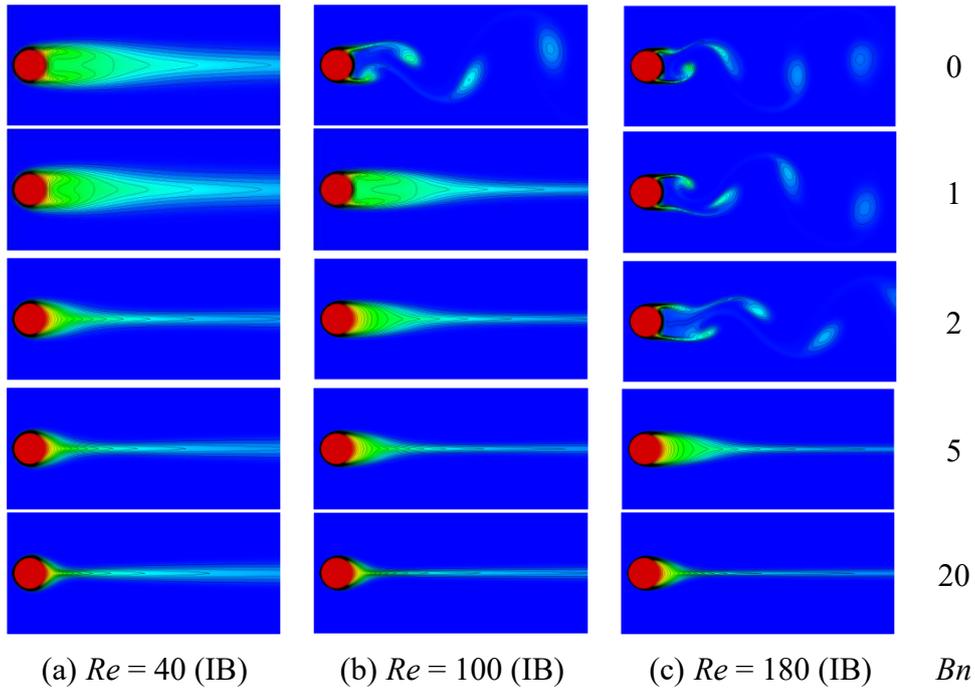

|  |  |  | Bn |
|---|---|---|---|
| (a) $Re$ = 40 (IB) | (b) $Re$ = 100 (IB) | (c) $Re$ = 180 (IB) |  |

Fig. 16. Effect of $Bn$ on the dimensionless temperature distributions with $Re$ = (a) 40, (b) 100, and (c) 180 for the IB process at a fixed $Pr$ = 10.

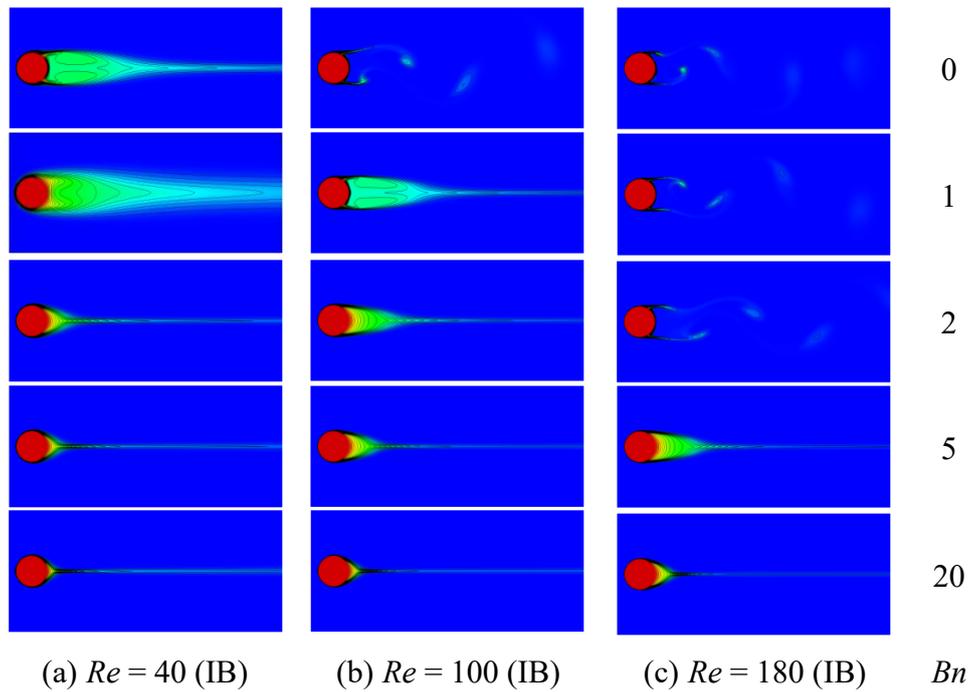

(a) $Re$ = 40 (IB)  (b) $Re$ = 100 (IB)  (c) $Re$ = 180 (IB)  $Bn$

Fig. 17. Effect of $Bn$ on the dimensionless temperature distributions with $Re$ = (a) 40, (b) 100, and (c) 180 for the IB process at a fixed $Pr$ = 100.

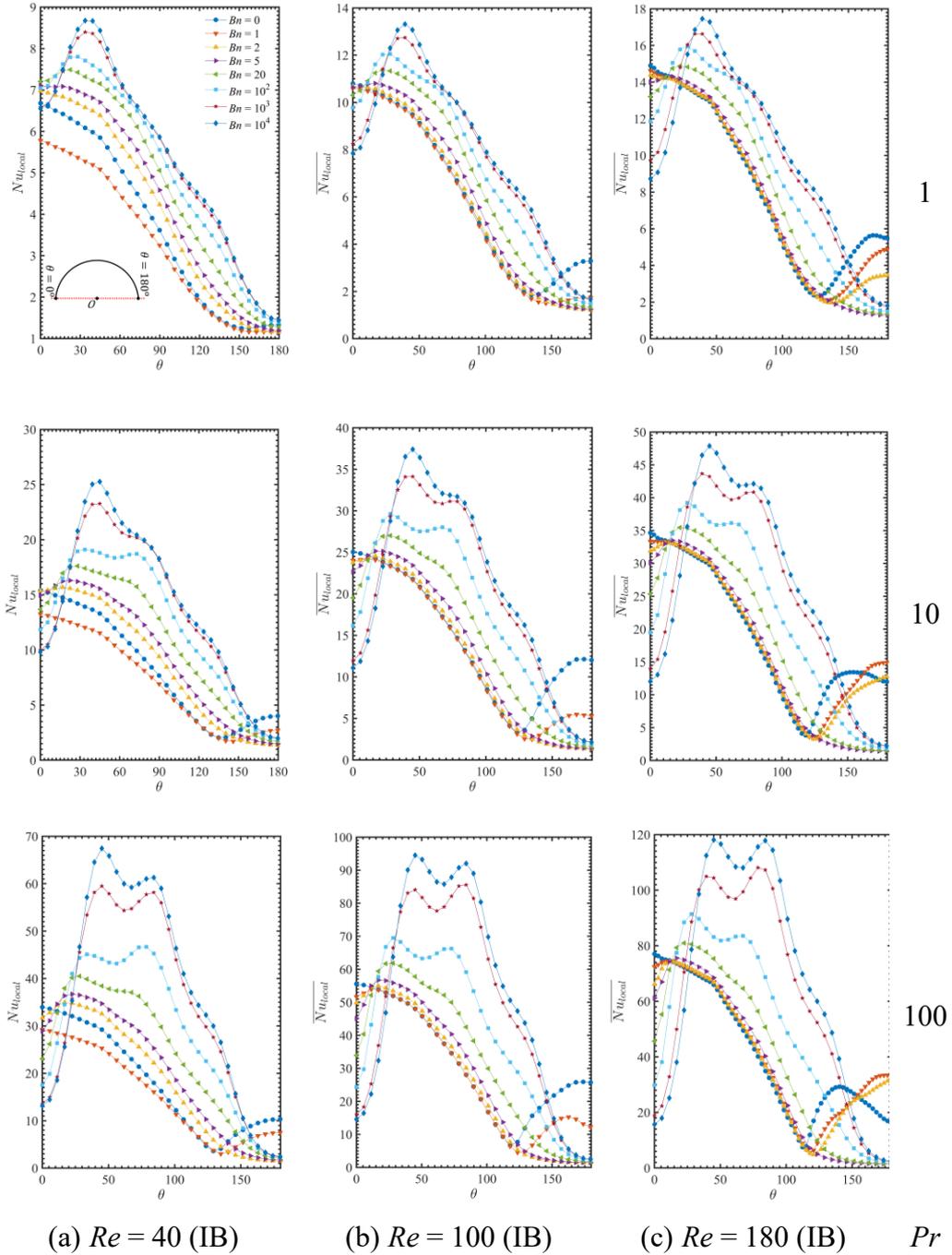

(a) $Re = 40$ (IB)    (b) $Re = 100$ (IB)    (c) $Re = 180$ (IB)    $Pr$

Fig. 18. The local Nusselt number profile along the cylinder surface at different $Bn$ and $Pr$ with $Re$ = (a) 40, (b) 100, and (c) 180 for the IB process.

The local Nusselt number $\overline{Nu_{local}}$ profiles along the cylinder surface at various $Bn$ and $Pr$ with $Re$ of 40, 100, and 180 for the IB process are illustrated in Fig. 18 ($Nu_{local}$ for the steady flow and $\overline{Nu_{local}}$ for the unsteady flow). The overall trend of $\overline{Nu_{local}}$ ($Nu_{local}$) with respect to $\theta$ displays a decreasing pattern. It is commonly observed that convective heat transfer along the upstream surface of the cylinder is typically more pronounced than that along the downstream surface. At low $Re$, $Bn$, and $Pr$, the location of the maximum local Nusselt number ($\overline{Nu_{local}}_{max}$) may appears at $\theta$ = 0° or another location on the upstream surface. For example, in the case of ($Re$, $Bn$, $Pr$) = (40, 0, 1), $\overline{Nu_{local}}_{max}$ is identified at $\theta$ = 0°, corresponding to the front stagnation point of the cylinder. In the cases of higher $Pr$ and $Bn$, two peaks may appear along the upstream surface, e.g. the case of ($Re$, $Bn$, $Pr$) = (180, $10^4$, 100). Notably, when $Bn$ is low, an extra-peak may appear near the rear part of the cylinder. The extra-peak locates around $\theta \approx 140.4°$ for the case of ($Re$, $Bn$, $Pr$) = (180, 0, 100).

Fig. 18 shows that, for a fixed $Re$ and $Bn$, the $\overline{Nu_{local}} \sim \theta$ curve shifts upwards and $\overline{Nu_{local}}_{max}$ increases with increasing $Pr$. This behavior is similar to that observed for a fixed $Re$ and $Pr$ with increasing $Bn$ (except for the extra-peak). For example, at ($Re$, $Bn$, $Pr$) = (40, $10^4$, 1), $\overline{Nu_{local}}_{max}$ is 8.664 and occurs at $\theta$ = 36.57° and no extra-peak point occurs near $\theta$ = 180°. At ($Re$, $Bn$, $Pr$) = (40, $10^4$, 10), $\overline{Nu_{local}}_{max}$ occurs at $\theta \approx 45°$. At ($Re$, $Bn$, $Pr$) = (40, $10^4$, 100), two peaks locate over the range from $\theta \approx 45°$ to $\theta \approx 85°$, and an extra-peak point occurs near $\theta$ = 180°. These results indicate that the extra-peak point near $\theta$ = 180° is more likely to appear at a higher $Pr$, highlighting the influence of $Pr$ on heat transfer around the cylinder.

For a fixed $Re$ and $Pr$, at a low $Bn$, such as $Bn$ = 1, with the increase of $Bn$, the $\overline{Nu_{local}}$ around the rear point $\theta$ = 180° decreases obviously while the extra-peak point near $\theta$ = 180° gradually disappears. This phenomenon correlates with the reduction of shear strain rate in the recirculation wake discussed in Figs. 9-12. The $\overline{Nu_{local}}$ profile near the front stagnation point of the cylinder ($\theta$ = 0°) mildly varies with $\theta$ at a low $Bn$. However, once $Bn$ surpasses a critical threshold, $\overline{Nu_{local}}$ begins to increases rapidly with $\theta$, which contributes to the shear strain rate enhancement in the boundary layer discussed in Figs. 9-12.

For a fixed $Bn$ and $Pr$, as $Re$ increases, both $\overline{Nu_{local}}_{max}$ and the $\overline{Nu_{local}} \sim \theta$ curve move upwards. For example, in the comparison between the cases of ($Re$, $Bn$, $Pr$)

= (40, 10$^4$, 1) and (*Re*, *Bn*, *Pr*) = (180, 10$^4$, 100), $\overline{Nu_{local}}_{max}$ increases from 8.664 to 118.1. However, the corresponding location $\theta_{max}$ for $\overline{Nu_{local}}_{max}$ does not exhibit significant changes with an increase in *Re*. Specifically, at (*Re*, *Bn*, *Pr*) = (40, 10$^4$, 1), $\overline{Nu_{local}}_{max}$ is observed at $\theta \approx 36.57°$; whereas at (*Re*, *Bn*, *Pr*) = (180, 10$^4$, 1), $\overline{Nu_{local}}_{max}$ occurs at $\theta \approx 39.38°$.

Generally, *Re* does not have a pronounced effect on the number of the peaks on the $\overline{Nu_{local}} \sim \theta$ curve. For example, at (*Re*, *Bn*, *Pr*) = (40, 10$^4$, 100) and (*Re*, *Bn*, *Pr*) = (180, 10$^4$, 100), two peaks exist on the $\overline{Nu_{local}} \sim \theta$ curve for both cases. However, *Re* may affect the extra-peak near $\theta = 180°$ when *Re* is beyond a certain threshold, e.g., the extra-peak is observed at (*Re*, *Bn*, *Pr*) = (180, 1, 1) but does not appear at (*Re*, *Bn*, *Pr*) = (40, 1, 1).

In general, with an increase in *Re*, *Pr*, or *Bn*, the $\overline{Nu_{local}} \sim \theta$ curve shifts upwards overall, accompanied by an increase in $\overline{Nu_{local}}_{max}$. With a rise in *Pr* or *Bn*, $\overline{Nu_{local}}_{max}$ moves toward the rear stagnation point of the cylinder. Furthermore, as *Re* and *Pr* increase while *Bn* decreases, the extra-pick near $\theta = 180°$ is more likely to emerge.

Table 5. The average Nusselt number on the cylinder surface in Bingham plastic fluids at *Pr* = 1.

| Re | Bn | Nu | |
|---|---|---|---|
| | | Nirmalkar & Chhabra[4] | Present |
| 20 | 10 | 3.5262 | 3.5097 |
| | 10$^4$ | 4.1917 | 4.0994 |

The comparison between our simulation results and those obtained by Nirmalkar & Chhabra[4] on the overall Nusselt number along the cylinder (*Nu*) for the steady flow of Bingham plastic fluids is listed in Table 5. The relative error between our simulation data and the findings of Nirmalkar & Chhabra[4] is less than 3%, indicating good agreement between the two results. An empirical formula for *Nu* based on (*Pr*, *Re*) proposed by Nirmalkar & Chhabra[4] is as follows,

$$Nu = 2.37 Re^{1/3} Pr^{1/3}. \tag{36}$$

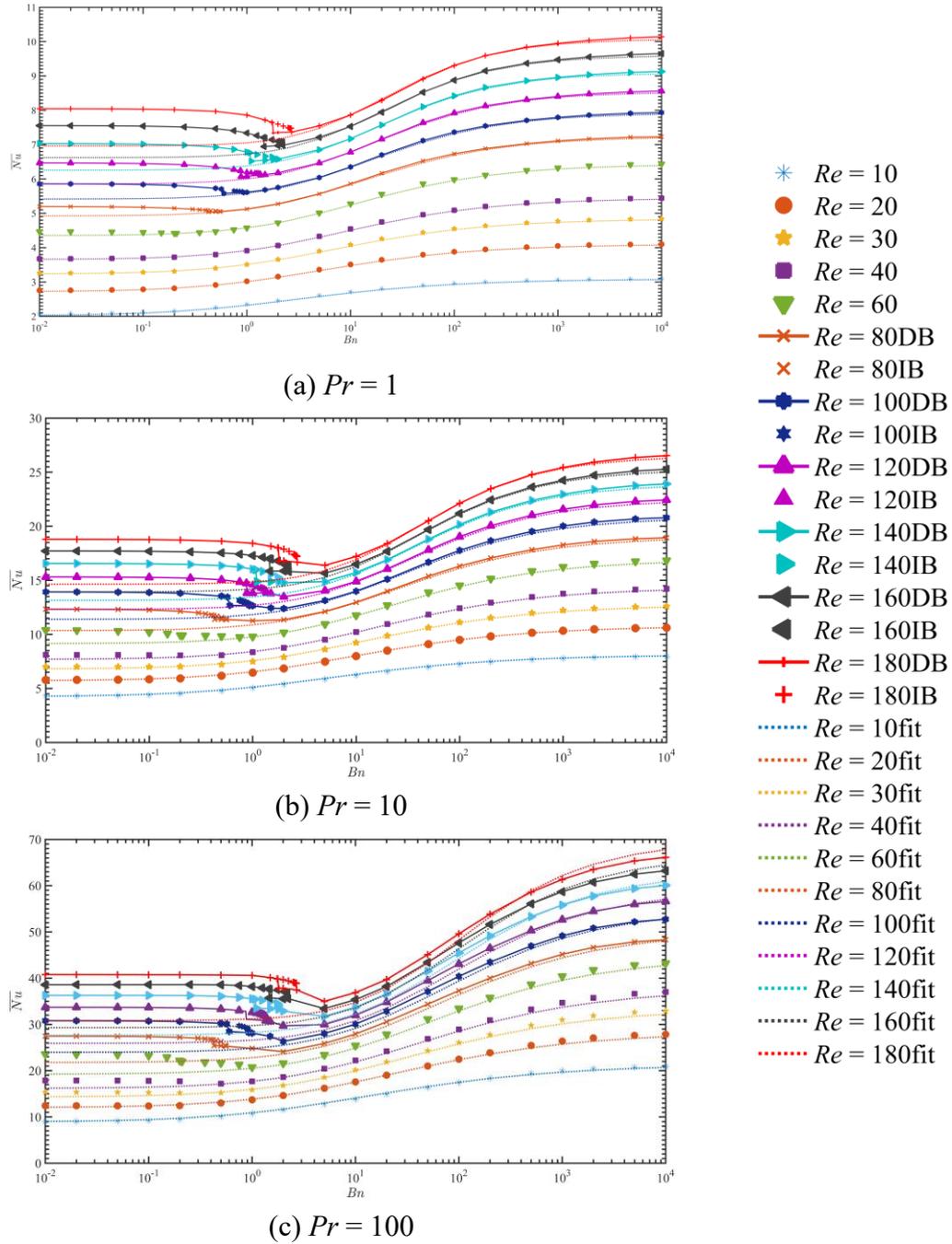

(a) *Pr* = 1

(b) *Pr* = 10

(c) *Pr* = 100

Fig. 19. Variation of $\overline{Nu}$ with *Bn* for different *Re* at *Pr* = (a) 1, (b) 10, and (c) 100. The dotted line in the figure indicates the fitting results with Eq. (37).

Variation of $\overline{Nu}$ with $Bn$ at different $Re$ and $Pr$ are depicted in Fig. 19. Here, $\overline{Nu}$ denotes the time-averaged Nusselt number along the cylinder surface after the flow reaches statistical stationary state for the unsteady flow while denotes the overall Nusselt number along the cylinder surface after the flow reaches the steady state for the steady flow. For all the cases with the same ($Re$, $Bn$), a higher $Pr$ corresponds to a higher $\overline{Nu}$, aligning with the trend described in Eq. (36). However, employing an exponential function of $Pr$ to describe $\overline{Nu}$ over the whole parameter space in this simulation is challenging. When $Re$ does not exceed 40, $\overline{Nu}$ monotonously increases with $Bn$ for a fixed $Re$ and $Pr$. The augmentation of $\overline{Nu}$ can be attributed to a thinner momentum boundary layer and a sharper shear strain rate are in the boundary layer discussed in Figs. 9-12. Conversely, for $Re = 60$ or higher and $Bn$ exceeds a threshold, $\overline{Nu}$ decreases with increasing $Bn$. For example, for $Re = 180$ (IB) and $Pr = 1$, $\overline{Nu}$ is 8.0468 at $Bn = 0$, and 7.3652 at $Bn = 2.7$. This reduction in $\overline{Nu}$ arises from the reduction in $\overline{Nu_{local}}$ near the rear of the cylinder while no obvious change in $\overline{Nu_{local}}$ along other parts of cylinder as described in Fig. 18. The essence of this reduction is the reduced shear strain rate behind the cylinder as shown in Figs. 9-12.

It is worth pointing out that near the two transitional points $Bn_{cI}$ and $Bn_{cD}$, the sudden change of flow field fluctuation (as shown in Fig. 13a) leads to abrupt variations in $\overline{Nu}$ with $Bn$. For example, at $Re = 180$ (IB) and $Pr = 1$, $\overline{Nu}$ is 7.4749 at $Bn = 2.68$, and $Nu$ is 7.3652 at $Bn = 2.7$. Similarly, at $Re = 180$ (DB) and $Pr = 1$, $Nu$ is 5.6026 at $Bn = 0.93$, and $\overline{Nu}$ is 5.6161 at $Bn = 0.92$. Due to the subcritical bifurcation in flow transition, the $\overline{Nu} \sim Bn$ curve in the IB process and the DB process displays inconsistency within the $Bn_{cI}$ and $Bn_{cD}$ intervals when $Re \geq 60$.

Eq. (34) is deemed applicable within the parameters of $1 \leq Re \leq 40$, $1 \leq Pr \leq 100$, and $0 \leq Bn \leq 10^4$. Despite the absence of an explicit influence of $Bn$ on $Nu$ as indicated in Eq. (36), it is evident that $Bn$ exhibits a substantial relationship with $Nu$ in the context of our study. Our study indicates that $Bn$ may need to be modified in the below formula,

$$\overline{Nu} = \overline{Nu}_0 + \left(\overline{Nu}_\infty - \overline{Nu}_0\right)\left[1 + (\lambda Bn)^{-\frac{n-1}{2}}\right]^{-2}, \qquad (Bn \geq 2 \cdot Bn_c) \qquad (37)$$

where,

$$\overline{Nu}_0 = 0.75505 Re^{0.42779} Pr^{0.322915}, \tag{38a}$$

$$\overline{Nu}_\infty = 1.2012 Re^{0.40964} Pr^{0.42006}, \tag{38b}$$

$$n = 2.00299 - 0.03361\log(Pr) + Re[0.002835 - 0.0001028\log(Pr)], \tag{38c}$$

$$\lambda = [8.26512 - 1.16921\log(Pr)]Re^{-0.81331 - 0.01388\log(Pr)}, \tag{38d}$$

where $\overline{Nu}_0$ and $\overline{Nu}_\infty$ are the overall Nusselt numbers of the cylinder at the limit of $Bn \to 0$ (the Newtonian fluid) and $Bn \to \infty$ (the fully plastic fluid), respectively, $n$ is the power-law index, and $\lambda$ is a parameter. The second term in the right-hand side of Eq. (37) is the increment of overall Nusselt number on the cylinder in Bingham fluid compared with that in Newton fluid.

Interesting, Eq. (37) is similar to the Carreau-Yasuda-like non-Newtonian viscosity model[43]. The best fitting results by Eq. (37) are also displayed in Fig. 19 for comparison. When the flow is steady (the right parts of the curves for a large $Bn$ in Fig. 19), the error between the fitting data and the original data is less than 5%. The corresponding error may be relatively large when the flow is unsteady. As shown in Fig. 4, when $Re$ exceeds $Re_c$, i.e., the flow transits from steady to unsteady, the $\overline{Nu} \sim Re$ relationship shows obvious discontinuity. Thus, Eq. (38a) can provide a better prediction for $\overline{Nu}_0$ in a Newtonian fluid when the flow is in a steady state. However, when the flow becomes unsteady, the flow fluctuation would significantly affect heat transfer, which pose challenges in the predication $\overline{Nu}_0$. This, therefore, leads to a relatively poor fit for the left parts of curves in Fig. 19.

Variations of $\overline{Nu}_0$ and $\overline{Nu}_\infty$ with $Re$ at different $Pr$ are shown in Figs. 20(a) and 20(b), respectively. Both $\log(Nu_0)$ and $\log(Nu_\infty)$ show the linear relationship with $\log(Re)$ and $\log(Pr)$, as expressed by Eq. 38(a) and (b). Variations of $n$ and $\lambda$ with $Re$ at different $Pr$ are shown in Figs. 20(c) and 20(d), respectively. The curve trends indicate that these two parameters have clear physical meaning. Take the derivative of Eq. (37) as follows,

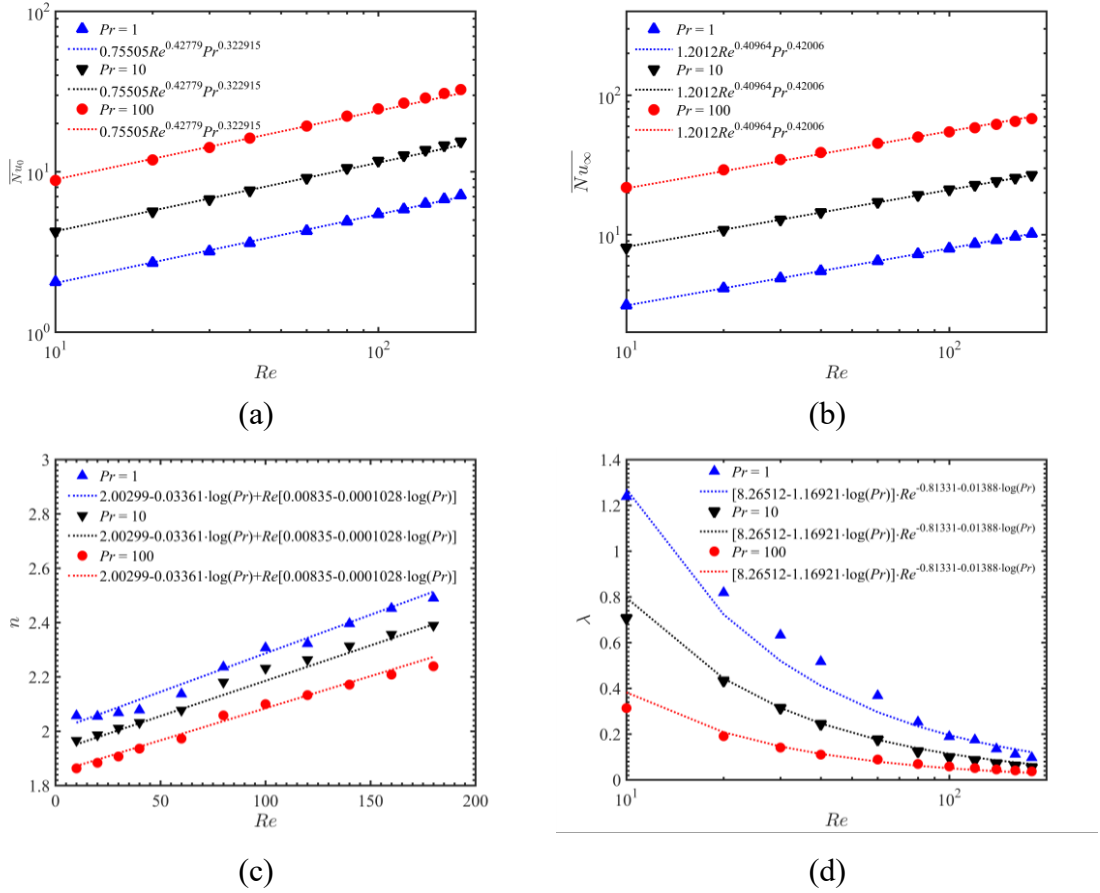

Fig. 20. Variations of (a) $\overline{Nu}_0$, (b) $\overline{Nu}_\infty$, (c) $n$, and (d) $\lambda$ with $Re$ at different $Pr$.

$$\frac{d \log(\overline{Nu}-\overline{Nu}_0)}{d \log(Bn)} = \frac{n-1}{\left[1+(\lambda \cdot Bn)^{\frac{n-1}{2}}\right]}. \tag{39}$$

When $\lambda Bn \ll 1$, we have

$$\frac{d \log(\overline{Nu}-\overline{Nu}_0)}{d \log(Bn)} = n-1. \tag{40}$$

Thus, $n-1$ represents the slope of the $(\overline{Nu} - \overline{Nu}_0) \sim Bn$ curve in the double logarithmic coordinate system when $\lambda Bn \ll 1$. $n$ is the linear combination of $Re$ and $\log(Pr)$ as shown in Fig. 20(c) and is large than 1 in the present parameter space. Eq. (37) could be written as the following form,

$$\log\left\{\left[\left(\frac{Nu-Nu_0}{Nu_\infty-Nu_0}\right)^2 - 1\right]^{-\frac{2}{n-1}}\right\} = \log(Bn) + \log(\lambda). \tag{41}$$

Eq. (41) indicates that $\lambda$ plays the role of curve transformation for the $\overline{Nu} \sim Bn$ relationship in the logarithmic coordinate system. $\lambda$ decays with $Re$ for a fixed $Pr$ as shown in Fig. 20(d). Correspondingly, the curves in Fig. 19 shift rightwards when $Re$ increases at a fixed $Pr$.

## 4. Conclusion

In various industrial applications, heat transfer in viscoplastic fluids, particularly Bingham plastic fluids, is critical, and optimizing the heat transfer process is essential for ensuring product quality and safety. Moreover, the limited understanding of the underlying mechanisms behind unsteady flow phenomena in Bingham plastic fluids, such as vortex shedding, needs to be studied more intensively. This study investigates the flow dynamics and heat transfer characteristics of a heated circular cylinder submerged in Bingham plastic fluids over wide ranges of parameter ranges with the plastic Reynolds number $10 \leq Re \leq 180$, the Prandtl number $1 \leq Pr \leq 100$, and the Bingham number $0 \leq Bn \leq 10^4$. Numerically results suggest that the flow fluctuation in the unsteady flow at a fixed $Re$ weakens gradually as $Bn$ increases. Beyond a critical value $Bn_c$, the flow becomes steady. This transition dissimilarity highlights the operational variance between the IB and DB processes. When $Re \geq 60$, the flow fluctuation near $Bn_{cI}$ or $Bn_{cD}$ undergoes a sudden change, reflected by sharp variation in $C_{lrms}$ with $Bn$. Consequently, sudden jumps occur near $Bn_{cI}$ and $Bn_{cD}$ in the $\overline{C_d}/\overline{C_d^{Newtonian}} \sim Bn$ curve and the $\overline{Nu}$ - $Bn$ curve. When $Re^* = Re/(1+Bn)$ is less than 0.5, $\overline{C_d}$ satisfies $\overline{C_d} = 24.84/Re^*$.

As $Re$, $Pr$ and $Bn$ increase, the $\overline{Nu_{local}} \sim \theta$ curve shifts upward, accompanied by an elevation in maximum local Nusselt number ($\overline{Nu_{local}}_{max}$). With increasing $Pr$ and $Bn$, $\overline{Nu_{local}}_{max}$ shifts towards the rear stagnation point of the cylinder. Generally, $Re$ does not have a pronounced effect on the number of the peaks on the $\overline{Nu_{local}} \sim \theta$ curve as well as the location of $\overline{Nu_{local}}_{max}$. Additionally, as $Re$ and $Pr$ increase and $Bn$ decreases, the extra-peak near $\theta = 180º$ becomes more prevalent. As $Bn$ increases, the variations in shear strain rate within the boundary layer exert a notable impact on the heat transfer characteristics of the cylinder. Furthermore, it is found that $\overline{Nu}$ and *Bn* fits well with the Carreau-Yasuda-like non-Newtonian viscosity model, especially for the steady flow.

## Acknowledgements

The authors would like to thank the financial support from the Department of Science and Technology of Guangdong Province (Grant No. 2023B1212060001), Shenzhen Science and Technology Innovation Commission (Grant No. JSGG20220831101400002), Guangdong Basic and Applied Basic Research Foundation (Grant No. 2022A1515011057) and the National Natural Science Foundation of China (NSFC, Grant Nos. 12172163, 12302361, 12071367, and 12002148). This work is supported by Center for Computational Science and Engineering of Southern University of Science and Technology.

## Data Availability Statement

The datasets used and/or analyzed during the current study are available from the corresponding author on reasonable request.

## Declaration of Competing Interest

The authors declare that there is no conflict of interest.